\documentclass[11pt]{article}
\usepackage{geometry}
\usepackage{amssymb}
\usepackage{amstext,enumerate}
\usepackage{bm}
\usepackage{graphics}
\usepackage{graphicx}
\usepackage{amsthm}
\usepackage{amsmath}
\usepackage{latexsym}
\usepackage{rotate}
\usepackage{lscape}
\usepackage{color}
\usepackage{epsfig,epsf,psfrag}
\usepackage{sectsty}
\usepackage{graphics}
\usepackage{epsfig}
\usepackage{multirow}
\usepackage{float}
\usepackage{graphicx}
\usepackage{stmaryrd}
\usepackage{amsmath}
\usepackage{url}
\usepackage{amsthm}
\usepackage{amssymb}
\usepackage{sectsty}
\usepackage{epsfig,natbib}
\usepackage{rotate}
\usepackage{lscape}
\usepackage{setspace}
\usepackage{subcaption}
\usepackage{float}
\usepackage{graphicx}
\usepackage{mathtools}
\usepackage{amstext,amssymb,amsfonts,amscd,color}
\usepackage{graphicx}
\usepackage{epsfig, natbib}
\usepackage{threeparttable}
\usepackage{cases,color}
\usepackage{multirow}
\usepackage{verbatim}
\usepackage{booktabs}
\usepackage{multicol}
\usepackage{bbm}
\usepackage{bm}
\usepackage{subcaption}
\usepackage{comment}

\include{mathdefn2}
\usepackage{mathrsfs}
\usepackage{float}

\usepackage{xparse}
\usepackage{tikz}
\usetikzlibrary{plotmarks}
\usetikzlibrary{matrix}
\usetikzlibrary{positioning}
\usepackage{pgfplots}
\pgfplotsset{compat=1.7}
\usetikzlibrary{matrix,backgrounds, arrows.meta}
\pgfdeclarelayer{myback}
\pgfsetlayers{myback,background,main}
\usetikzlibrary{decorations.pathreplacing,angles,quotes}
\tikzset{mycolor/.style = {dashed,rounded corners,line width=1bp,color=#1}}%
\tikzset{myfillcolor/.style = {draw,fill=#1}}%
\tikzset{
	declare function={
		normcdf(\x,\m,\s)=1/(1 + exp(-0.07056*((\x-\m)/\s)^3 - 1.5976*(\x-\m)/\s));
	}
}

\sectionfont{\centering\bf\sf\normalsize}
\subsectionfont{\sf\normalsize}

\renewcommand{\baselinestretch}{1.6} 
\newcommand{\single}{\renewcommand{\baselinestretch}{1.2}\normalsize}
\newcommand{\double}{\renewcommand{\baselinestretch}{1.63}\normalsize}

  \renewenvironment{thebibliography}[1]{
    \begin{oldthebibliography}{#1}
      \setlength{\parskip}{0ex}
      \setlength{\itemsep}{0ex}
  }
  {
    \end{oldthebibliography}
  }

\newcommand{\bea}{\begin{eqnarray*}}
\newcommand{\eea}{\end{eqnarray*}}
\newcommand{\be}{\begin{eqnarray}}
\newcommand{\ee}{\end{eqnarray}}
\newcommand{\ed}{\end{document}}

\newcommand{\btab}{\begin{tabular}}
\newcommand{\etab}{\end{tabular}}

\newcommand{\bi}{\begin{itemize}}
\newcommand{\ei}{\end{itemize}}
\newcommand{\bfi}{\begin{figure}}
\newcommand{\efi}{\end{figure}}
\newcommand{\ben}{\begin{enumerate}}
\newcommand{\een}{\end{enumerate}}
\newcommand{\bay}{\begin{array}}
\newcommand{\eay}{\end{array}}

\newcommand{\rs}{\vspace{-.2cm}}

\def\vs{\vspace{.5cm}}
\def\vvs{\vspace{.15cm}}

\definecolor{DarkBlue}{rgb}{0,.08,.45}
\definecolor{DarkRed}{rgb}{.7,0,.4}

\def\hg #1 {\texcolor{cyan}{{\it Hans:}   #1}}

\def\bco{\iffalse}

\def\cp{\citep}

\def\rs{\vspace{-.1in}}

\newcommand{\no}{\noindent}
\newcommand{\bc}{\begin{center}}
\newcommand{\ec}{\end{center}}

\newcommand{\bsp}{\begin{split}}
\newcommand{\esp}{\end{split}}
\newcommand{\bdes}{\begin{description}}
\newcommand{\edes}{\end{description}}
\newcommand{\bass}{\begin{assumption}}
\newcommand{\eass}{\end{assumption}}
\newcommand{\bthm}{\begin{theorem}}
\newcommand{\ethm}{\end{theorem}}
\newcommand{\blem}{\begin{lemma}}
\newcommand{\elem}{\end{lemma}}

\def\bco{\iffalse}

\def\cp{\citep}

\DeclareMathOperator*{\argmin}{argmin}

\def\HS{\mathcal{S}^\infty}

\def\H{\mathcal{H}}
\def\S{\mathcal{S}}
\def\real{\mathbb{R}}

\newcommand{\sm}{}
\bibpunct{(}{)}{;}{a}{}{,}

\begin{document}
\thispagestyle{empty} \single \bc {\bf \sc \Large Spherical  Autoregressive Models, With Application to  Distributional and Compositional Time Series} 
\vspace{0.15in}\\
Changbo Zhu and  Hans-Georg M\"uller \\
Department of Statistics, University of California, Davis \\
Davis, CA 95616 USA \ec \centerline{7 October 2021}

\vspace{0.1in} \thispagestyle{empty}
\bc{\bf \sf ABSTRACT} \ec \vspace{-.1in} \no 
\setstretch{1} 
We introduce a new class of autoregressive models for spherical time series, where the dimension of the spheres on which the observations of the time series are situated may be  finite-dimensional or infinite-dimensional as in  the case of a general Hilbert sphere. Spherical time series arise in various settings. We focus here on distributional and compositional time series. Applying a square root transformation to the densities of the observations of a distributional time series  maps the distributional observations  to  the Hilbert sphere, equipped with the Fisher-Rao metric. Likewise, applying a square root transformation to the components of the observations of a compositional time series maps the compositional observations to a finite-dimensional sphere,   equipped with the geodesic metric on spheres.  The challenge in modeling such time series lies in the intrinsic non-linearity of spheres and Hilbert spheres, where conventional arithmetic operations such as addition or scalar multiplication are  no longer available. To address this difficulty, we consider rotation operators to map observations on the sphere. Specifically, we introduce a  class of  skew-symmetric operator such that the associated exponential operators are rotation operators  that for each given pair of points on the sphere map one of the points to the other one. 
We exploit the fact that the  space of skew-symmetric operators is Hilbertian to develop  autoregressive modeling of geometric differences that correspond to rotations of  spherical and distributional time series.   Differences expressed in terms of rotations  can be taken between the Fr\'{e}chet mean and the observations  or between consecutive observations of the time series.  We derive theoretical properties of the ensuing  autoregressive models and showcase these approaches with several motivating data. These include a time series of  yearly observations of bivariate distributions  of the minimum/maximum temperatures for  a period of  120 days during  each summer for the years  1990-2018 
at Los Angeles (LAX) and John F. Kennedy (JFK) international  airports.  A second data application concerns a   compositional time series with annual observations of compositions of energy sources for power generation in the U.S..\\

\no {KEY WORDS:\quad Distributional Data, Compositional Data, Hilbert Sphere, Fisher-Rao Metric, Geodesics, Skew-Symmetric Operators, Rotation Operators, Random Objects}.
\thispagestyle{empty} \vfill
\noindent \vspace{-.2cm}\rule{\textwidth}{0.5pt}\\
{\small Research supported in part by NIH Echo and NSF DMS-2014626.}

\newpage
\pagenumbering{arabic} \setcounter{page}{1} \double

\bc {\bf \sf \Large 1.\quad Introduction}\sm \ec \rs

Modern day  data analysts  increasingly encounter  complex data types where data  are no longer traditional vectors, and furthermore are not  situated in a linear space such as a Hilbert space. Such non-Euclidean data may also be encountered in the form of a  time series. At this point,  the methodology available for the analysis of such data is quite limited. An exception are recent efforts to  develop models for distributional time series in the context of the rapidly evolving field of distributional data analysis (DDA) \cp{pete:21}. 
A simple approach for distributional time series is to  represent distributions by square integrable functions via  the log quantile density transformation or a  similar transformation \cp{mull:16:1}; a downside is that such transformations may lead to large metric distortions. The distributional time series is then transformed to a functional time series, which have been well investigated \cp{bosq:00}. Geometric approaches that are based on constructing tangent bundles on the Wasserstein manifold have recently been shown to provide better predictions for autoregressive models \cp{mull:21:4,zhan:21}, while an autoregressive model that is intrinsic to the Wasserstein manifold  can be based on  a recently developed transport algebra \cp{mull:21:3}. It bears emphasizing that all these methods are limited to the case of distributional time series composed of one-dimensional distributions. 

Modeling distributional time series has been limited to the case of sequences of one-dimensional distributions, as the challenges of characterizing optimal transport as well as the Wasserstein manifold and its parallel transport for the case of multivariate distributions are formidable,  in addition to numerical difficulties, and viable transformations have not yet been developed.  For both multivariate and one-dimensional distributions the Fisher-Rao metric provides an alternative to the popular Wasserstein metric that is easily to work with both numerically and theoretically,  irrespective of the dimension of the distributions. This distributional metric is characterized by its invariance under diffeomorphisms and the ease of explicitly computing geodesics in the space of distributions with smooth densities equipped with this metric \cp{frie:91,baue:16}. 
Of special interest for statistical applications is that the Fisher-Rao metric  can be easily extended to multivariate distributions, and neither analysis nor numerical implementations face difficulties in the multivariate case, and the  geodesics in distribution space are always well-defined irrespective of the dimension. 

We focus here on time series data with observations that  reside naturally or can be equivalently represented as points on a sphere $\mathcal{S} = \{ x \in \mathcal{H}  :  \| x \|_{\mathcal{H}} =1  \}$, where $\mathcal{H}$ is a real separable Hilbert space with inner product $ \langle \cdot , \cdot \rangle_{\mathcal{H}} $ and norm $\| x \|_{\mathcal{H}} := \sqrt{  \langle x , x \rangle_{\mathcal{H}}}$. The sphere $\mathcal{S}$  can be finite-dimensional in which case we denote it by $\mathcal{S}^d$ if $\H=\real^{d+1}$ or infinite-dimensional when 
$\H=L^2$ or any isomorphic space and in this case we refer to it as the Hilbert sphere and denote it by $\mathcal{S}^{\infty}$. 
Our focus on spherical time series is motivated by the convenience of incorporating different data types such as compositional data, directional data and distributional data. 

Compositional data take values in the simplex 
\begin{align*}
    \mathcal{C}^d = \left\{ \mathbf{z} = (z_1, z_2, \cdots, z_d)^T \} \in \mathbb{R}^d \; \left| \; z_i \ge  0 \text{ for all } i =1,2, \cdots, d \text{ and } \sum_{i=1}z_i = \kappa \right. \right\}, 
\end{align*}
where $\kappa >0$ is a constant with default value  $\kappa=1$, in which case these data are non-negative proportions that sum to 1.    By applying the point-wise square root ratio (psr) transformation $\text{psr}: \mathcal{C}^d \rightarrow \mathcal{S}^{d-1}$, defined as   
\begin{align}
\label{eq:psr}
    \text{psr}(\mathbf{z}) = \left(\sqrt{z_1/\kappa}, \sqrt{z_2/\kappa}, \cdots, \sqrt{z_d/\kappa} \right)^T, 
\end{align} 
$\mathcal{C}^d$ can be mapped into a  subset of $ \mathcal{S}^{d-1}$. This maps compositional time series to finite-dimensional spherical time series \cp{scea:11,mull:18:4}.  Examples of compositional time series are common and include for example repeated election cycles  when there are several parties and the compositions correspond to the vote shares of each party; or color preferences of car buyers that change from year to year, reflected in the percentage of cars sold in a specific color.   In section 5.2, we illustrate the proposed sperical autoregressive models (SAR) with  compositional time series that correspond to the annually recorded  proportions of electricity generated from different energy sources in the U.S.,  where energy sources  include coal, natural gas or nuclear and renewable sources. The composition of energy sources has a major impact on the  carbon dioxide (CO2) emissions that accrue from electricity generation over time.

Data that can be represented by locations  on finite-dimensional spheres are ubiquitous and are not limited to compositional data but also include directional data such as wind directions. For example, the study of  ocean surface wind over time is important in determining the spread of aerial organisms \citep{felicisimo2008ocean}. Sequences of hourly or daily recorded wind directions are naturally represented as a spherical time series with observations  in  $\mathcal{S}^2$. Another application of $\mathcal{S}^d$-valued time series are vector time series, where the vector observations can  be expressed in polar coordinates and then form a spherical time series and a scalar time series where the latter corresponds to the length of the vector at time $t$.  In some cases the length of the vector may not be relevant if one is primarily interested in the association between the vector components as reflected by the direction of the vector. Then only  the sequence of directions of the vector components matters and if the original vector data have dimension $d+1$, the directions are represented on $\mathcal{S}^d$ and again one has a spherical time series.

For any density function $f: \mathbb{R}^D \rightarrow \mathbb{R}$, where  $f$ satisfies $f\geq 0$ and $\int_{\mathbb{R}^D} f(x) dx = 1$, we define the functional point-wise square root transformation (fpsr) as 
\begin{align*}
\text{fpsr} (f) = g, \text{ where }g(\mathbf{z})= \sqrt{f(\mathbf{z})} \text{ for all } \mathbf{z}  \in \mathbb{R}^D.
\end{align*}
 Using $\text{fpsr}$,  distributional data  correspond to the  elements of a segment of the Hilbert sphere $\HS$ equipped with the Fisher-Rao metric \cp{dai:21}  and distributional time series then are accordingly represented  as $\HS$-valued time series. For two-dimensional distributions,  of  daily  maximum and minimum temperatures recorded for 24 hours over  the summer months at airports in the U.S. Considering these two-dimensional distributions over successive years then forms a time series with $\HS$-valued observations. These time series are of interest for assessing the effects of climate change and the  risks and costs associated with  rising  temperatures.

Time series analysis for Euclidean vector data   is a well-established field and both parametric and non-parametric methods have been developed.  \citep{fan2017elements, fan2008nonlinear}.
For functional or  Hilbert-space valued time series  linear and autoregressive process models have also been well studied, starting with  \cite{bosq:00}. In contrast to these developments,  there is so far very little work on time series with random objects, i.e., random variables  in general metric spaces \cp{mull:16:9}. Even for   the special case of spherical time series  the literature is scarce. An interesting approach is spherical regression for the non-time series case when one has i.i.d. data with predictors and responses located in  $\mathcal{S}=\mathcal{S}^d$ \citep{Tedchang, tedchangerror, kim, macro2019jasa}, where the key ingredient is  a rotation matrix in the set of orthogonal matrices $\text{SO}(d+1)$ that rotates the predictor to the response. In addition, \cite{down} and \cite{michael} introduced some more general families of transformations and \cite{tianxicai} investigated settings where predictors and responses might have mismatches. However, all these methods are established under the i.i.d. regression setting and limited to  the finite-dimensional case ($d<\infty$); furthermore,  they accommodate only one predictor, while for autoregressive modeling one needs to accommodate the joint action of predictors from multiple lags. 

The main challenge of modeling time series in non-linear spaces such as $\mathcal{S}$ is that conventional operations like addition and subtraction are not available. This lack of algebraic operations imposes  a fundamental limitation for autoregressive modeling. To  overcome  the challenge of  non-linearity for the case of spherical time series, we utilize the geometric structure of $\mathcal{S}$. The geodesic distance on $\mathcal{S}$ is defined as $d(x_1, x_2) = \text{arccos}(\langle x_1, x_2 \rangle) $ for any $x_1, x_2 \in \mathcal{S}$. Geodesics are locally length-minimizing paths between points that are well-defined in geodesic metric spaces, where the length of a geodesic path between two points coincides with the  distance of the points \cp{bura:01}.  The geodesics of spheres correspond to great circles. The key idea for the modeling of spherical autoregressive (SAR) time series is that the geodesic between two points $x_1, x_2 \in \mathcal{S}$ can be written as $\gamma(a) = \exp(a L)$, where $a \in [0,1]$ and $L:\mathcal{H} \rightarrow \mathcal{H}$ is a skew-symmetric operator. 
We then relate the spherical difference between $x_1$ and $x_2$ to the operator  $L$, which is a linear operator.  This makes it possible to model the differenced times series in the linear space of skew-symmetric operators. 

We study two versions of autoregressive models for spherical time series. In the basic SAR model the autoregressive model is based on  the spherical equivalent of differences between the observations and the overall Fr\'{e}chet mean as arguments. A second model, referred to as    DSAR, is based  on the spherical differences between consecutive observations.  These  models can be applied for autoregressive modeling on spheres  $\mathcal{S}$ of finite or infinite  dimension and their implementation is computationally efficient.

The rest of the paper is organized as follows. In Section 2, we introduce the rotation and log rotation operators in Hilbert spaces and present a key relationship between rotations and skew-symmetric operators.  Methodology and theory  are  in Section 3, which contains the main results.  Estimation and prediction are studied in Section 4. We report  results for data applications to distributional and compositional time series in Section 5, which is followed by a discussion section in Section 6. \vspace{.5cm}

\bc {\bf \sf \Large 2.\quad Rotations  and Skew-Symmetric  Operators}\sm \ec \rs \vvs

Let $\mathcal{H}$ be a real separable Hilbert space with inner product $ \langle \cdot , \cdot \rangle_{\mathcal{H}} $ and norm $\| x \|_{\mathcal{H}} := \sqrt{  \langle x , x \rangle_{\mathcal{H}}}$. The Hilbert sphere $\mathcal{S}$ is a subset of $\mathcal{H}$ whose elements have norm 1, i.e.,  $\mathcal{S} = \{ x \in \mathcal{H}  :  \| x \|_{\mathcal{H}} =1  \}$. Given a set of points $\{ x_1, x_2, \cdots, x_m \} \subset \mathcal{S}$, let 
$\text{span}\{ x_1, x_2, \cdots, x_m \} = \{ a_1x_1 + a_2x_2 + \cdots + a_m x_m : \, a_1, a_2, \cdots, a_m \in \mathbb{R} \} \subset \mathcal{H}$ 
denote the $m$-dimensional subspace of $\mathcal{H}$  spanned by $x_1, x_2, \cdots, x_m$.  The set of bounded linear operators on $\mathcal{H}$ is denoted as $\mathscr{B}(\mathcal{H})$ 
and an operator $Q \in \mathscr{B}(\mathcal{H}) $ is skew-symmetric if 
\begin{align*}
    \langle Qx, y \rangle + \langle x, Qy \rangle = 0 \text{ for all }x, y \in \mathcal{H}.
\end{align*}
For any bounded linear operator $L:\mathcal{H} \rightarrow \mathcal{H}$, define its exponential as $\text{exp}(L) := \sum_{l=0}^{\infty} L^l/l!$.
An orthogonal operator $O \in \mathscr{B}(\mathcal{H})$ is a rotation operator if and only if there exists a skew-symmetric operator $Q$ such that $ O = \text{exp}(Q)$  \citep{monroe1932}. 

For each skew-symmetric operator $Q$ there is a unique  rotation operator exp$(Q)$.
Let $\mathscr{R}(\mathcal{H})$ and $ \mathscr{S}(\mathcal{H}) $ be the set of rotation operators and skew-symmetric operators respectively, then  by  definition $ \mathscr{R}(\mathcal{H}) = \exp ( \mathscr{S}(\mathcal{H}) ) $. 
If $\{ e_1, e_2, \cdots \}$ is an orthonormal basis of $\mathcal{H}$, then $\mathscr{S}(\mathcal{H})$ admit the following orthonormal basis
\begin{align*}
    \mathscr{S}(\mathcal{H}) = \text{span}\left\lbrace e_i \otimes e_j - e_j \otimes e_i: i,j = 1, 2, \cdots  \right\rbrace \subset \mathcal{H} \otimes \mathcal{H},
\end{align*}
where $\mathcal{H} \otimes \mathcal{H}$ is the tensor product of the Hilbert space $\mathcal{H}$ with itself and is also a Hilbert space with inner product
\begin{align*}
\langle x_1 \otimes y_1, x_2 \otimes y_2 \rangle_{\mathcal{H} \otimes \mathcal{H}} = \langle x_1 , x_2  \rangle_{ \mathcal{H}} \langle y_1 , y_2  \rangle_{ \mathcal{H}},
\end{align*}
with $x_1, x_2, y_1, y_2 \in \mathcal{H}$; the inner product $\langle \cdot, \cdot \rangle_{\mathcal{H} \otimes \mathcal{H}} $ can be extended to any element in $\mathcal{H} \otimes \mathcal{H}$ by linearity. Observing that 
$\mathscr{S}(\mathcal{H}) $ is a closed subspace of $ \mathcal{H} \otimes \mathcal{H} $ with respect to $\langle \cdot, \cdot \rangle_{\mathcal{H} \otimes \mathcal{H}} $, $ \mathscr{S}(\mathcal{H}) $ is seen to be a complete separable Hilbert space. 

Given two points $x_1, x_2 \in \mathcal{H}$ such that $x_1 \neq x_2$ and $x_1 \neq -x_2$,
the proposed methodology relies on rotation operators that provide a rotation on $\S$  within the two dimensional subspace $\text{span}\{ x_1, x_2 \}$.\vs

\no {\bf Theorem 1.} {\it Set $ u_1 = x_1 \text{ and } u_2 = (x_2 - \langle x_2, u_1 \rangle u_1) / \| x_2 - \langle x_2, u_1 \rangle u_1 \|_{\mathcal{H}} $.  Let $I$ be the identity operator and  $Q := u_1 \otimes u_2 - u_2 \otimes u_1 \in \mathscr{S}(\mathcal{H})$. Then, given an angle $\vartheta \in [0, 2 \pi]$,
\begin{align}
\label{eq:main}
\exp( \vartheta Q) = I + \text{sin}(\vartheta) Q + (1 - \text{cos}(\vartheta))Q^2 
\end{align}
is a rotation operator that rotates counterclockwise within $\emph{\text{span}}\{ u_1, u_2 \}$ by $\vartheta$, i.e., 
\begin{itemize}
    \item For any $y_1, y_2 \in \mathcal{H}$, $\langle \exp( \vartheta Q)y_1, \exp( \vartheta Q)y_2 \rangle_{\mathcal{H}} = \langle y_1, y_2 \rangle_{\mathcal{H}}. $
    \item For any $x \in  \emph{\text{span}}\{ u_1, u_2 \} \cap \mathcal{S}$, $\emph{\text{arccos}} \langle \exp( \vartheta Q)x, x \rangle_{\mathcal{H}} = \vartheta .$
    \item For any $y \in \mathcal{H}$ perpendicular to $ \emph{\text{span}}\{ u_1, u_2 \} $, i.e.,  $\langle y, u_1 \rangle_{\mathcal{H}} = 0$ and $ \langle y, u_2 \rangle_{\mathcal{H}} = 0 $ , it holds that $ \exp( \vartheta Q) y= y $.
\end{itemize}
}

\begin{figure}[h]
\begin{center}
    \begin{tikzpicture}[scale=1.3,rotate=10]

  \def\r{3}
 
  \draw (0,0) node[circle, fill, inner sep=1] (orig) {};
  
  \draw (orig) circle (\r);
  \draw[fill = green!30, dashed] (orig) ellipse (\r{} and \r/3);

  \draw node [circle, fill, inner sep=1] (orig) {};

  \draw[dashed] (orig) -- (\r/3*1.45, -\r/5*1.45) node[circle, fill,inner sep=1.5, label=below:$\exp(\vartheta Q) x_1$] (phi) {}; 
  
  \draw[draw=cyan, text=cyan, ->, rotate=-10] (orig) -- ++(0, \r*1.2) node[right] (y) {z-axis};
  \draw[draw=cyan, text=cyan, ->, rotate=-10] (orig) -- ++(\r*1.2, 0) node[below] (x2) {x-axis};
  \draw[draw=cyan, text=cyan, ->, rotate=-10] (orig) -- ++(-\r/5*2, -\r/3*2) node[below] (x1) {y-axis};

  \draw[ thick,->] (orig) -- ++(-\r/5, -\r/3) node[circle, fill,inner sep=1.5,label=below:$x_1 (u_1)$] (x1) {};
  \draw[dashed] (orig) -- ++(\r/1.25, -\r/5) node[circle, fill,inner sep=1.5, label = below:$x_2$] (x3) {};
  \draw[thick, ->] (orig) -- ++(\r, 0) node[circle, fill,inner sep=1.5, label=right:$u_2$] (x2) {};
  
  \pic [draw=black, text=black, ->, "$\vartheta$"] {angle = x1--orig--phi};
\clip(0,0)--(-\r/5*2, -\r/3*2)--(\r/1.2*1.4, -\r/5*1.5)--cycle;
  
  \draw[color=blue, ultra thick] (0,0) circle [x radius=\r, y radius=\r/3];

\end{tikzpicture}
\end{center}
    \caption{Illustration of the rotation operator $\exp(\vartheta Q)$ when $\mathcal{H} = \mathbb{R}^3$. The green plane is the two-dimensional subspace spanned by $u_1$ and $u_2$. By construction, $u_1$, $u_2$ are orthogonal and the angle between them is $\pi/2$. Here $\exp(\vartheta Q)x_1$ is the location of the image of the  rotation operator $\exp (\vartheta Q)$ applied at  $x_1$ and  $\vartheta$ is the angle between $x_1$ and $\exp (\vartheta Q)x_1$. The blue line is the geodesic between $x_1$ and $x_2$ that is traced by the path $\gamma(a) := \exp (a \theta Q) x_1$, where $a \in [0,1]$ and $\theta = \text{arccos}(\langle x_1, x_2 \rangle)$. It can be easily seen that $\gamma(0)=x_1$ and $\gamma(1) = x_2$.}
    \label{fig:rotation}
\end{figure}
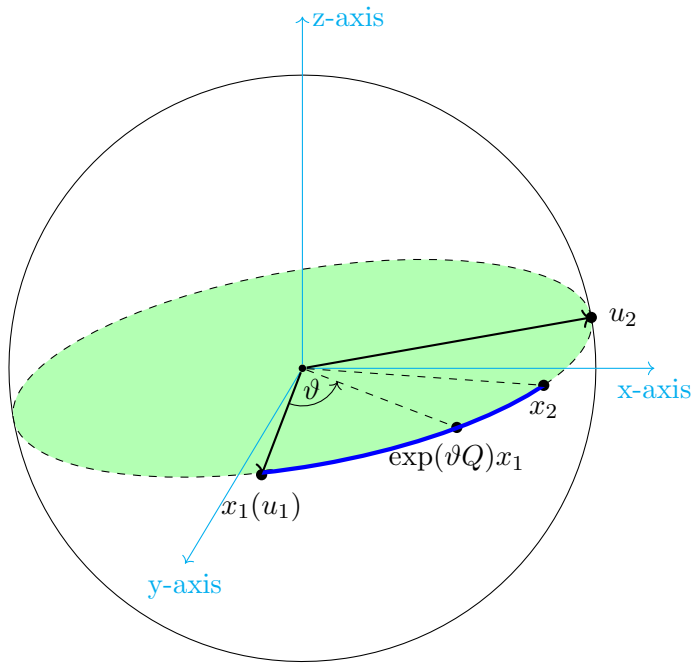

For  $\mathcal{H} = \mathbb{R}^3$,  Figure \ref{fig:rotation} provides  an illustration of the rotation operator $\text{exp}(\vartheta Q)$. We note that  \eqref{eq:main} reduces to the Rodrigues rotation formula in  this special case. For a rotation operator $\exp(\vartheta L)$ in higher dimensional Hilbert spaces such as $\mathbb{R}^d$ with $d>3$, where $L$ is an arbitrary skew-symmetric operator, the equality $\exp( \vartheta L) = I + \text{sin}(\vartheta) L + (1 - \text{cos}(\vartheta))L^2$  will not hold in general. That  \eqref{eq:main} is satisfied  for any separable space $\mathcal{H}$ is due to the fact that $\exp (\vartheta Q)$ is a special  rotation that only rotates within the two-dimensional subspace $\text{span}\{u_1, u_2\}$.


\bc {\bf \sf \Large 3.\quad Spherical Autoregressive Models}\sm \ec \rs

Based on the rotation operator introduced earlier, the geodesic  $\gamma:[0,1] \rightarrow \mathcal{S}$ between two points $x_1, x_2 \in \mathcal{S}$ can be traced  by rotating $x_1$ to $x_2$ within the two dimensional subspace spanned by $\{x_1, x_2\}$ around the origin, i.e.,
\begin{align*}
\gamma(a) = \exp(a\theta Q)x_1.
\end{align*}
where $a \in [0,1]$, $\theta = \text{arccosin}(\langle x_1, x_2 \rangle)$ is the angle between $x_1$ and $x_2$ and $Q$ is the same as in Theorem 1; see Figure \ref{fig:rotation} for a demonstration of $\gamma(a)$ when $\mathcal{H} = \mathbb{R}^3$. We then  utilize geodesics on $\S$  to arrive at a notion of difference between points on $\mathcal{S}$. Starting with the  Euclidean space $\mathbb{R}^d$ and considering two elements $w_1, w_2 \in \mathbb{R}^d$, the difference $V = w_2 - w_1$ can be interpreted as the optimal transport map that moves $w_1$ to $w_2$ and the geodesic between them is the straight line $r(a) = w_1 + a V$ where $a \in [0,1]$; see \cite{mull:21:3} for a similar extension of this idea to the Wasserstein space. On the other hand, not only $\exp(\theta Q)$ may be interpreted as the optimal transport map that moves $x_1$ to $x_2$ but also the geodesic can be constructed based on $\exp (\theta Q)$. This motivates  us to define the spherical difference between points $x_1$ and $x_2$ on $\S$,
\begin{align*}
   x_2 \ominus x_1 := \theta Q.
\end{align*}

Given a sequence of data points $x_1, x_2, \cdots, x_n \in \mathcal{S}$ with the same  Fr\'{e}chet mean $\mu_x$, i.e., $\mu_x := \argmin_{z \in \mathcal{S}} \mathbb{E} [ d_{\mathcal{S}}^2(z, x_t)]\text{ for all } t =1,2, \cdots, n$, we then construct a new series by taking  differences between the $x_t$ and the  Fr\'{e}chet mean $\mu_x$, 
\begin{align*}
\{ R_t := x_t \ominus \mu_x :  t =1,2, \cdots, n \} \subset \mathscr{S}(\mathcal{H}).
\end{align*} 
Assuming  that $\{ R_t \}$ is a stationary sequence \citep{bosq:00}, we propose the following spherical autoregressive (SAR) model of order $p$,
\begin{align} \label{model1}
R_t-\mu_{R} = \alpha_1 (R_{t-1} - \mu_{R}) + \cdots + \alpha_p (R_{t-p} - \mu_R) + \varepsilon_t \text{ where } R_t =  x_{t} \ominus \mu_x,
\end{align}
where  $\alpha_1, \cdots, \alpha_p \in \mathbb{R}$, $\mu_R = E[R_t]$ and  $\{ \varepsilon_t \} \subset \mathscr{S}(\mathcal{H})$ are i.i.d random innovations with mean 0. 

To elucidate the connection of this model with the previously studied spherical regression \citep{Tedchang, tedchangerror, kim, macro2019jasa}, which has not yet been extended to a time series framework and admits only one predictor, consider a regression setting with $x_t$ as single predictor and $y_t$ as response. In the above difference notation, this previously studied spherical  regression  can  be written  as $y_t \ominus x_t = R_0 + \varepsilon_t$. In the Euclidean space $\mathbb{R}^d$ this corresponds to an  intercept only regression model $ z_t - w_t = \beta_0 + \epsilon_t$, where $ z_t \in \mathbb{R}^d$ is the response, $w_t \in \mathbb{R}^d$ is the predictor, $\beta_0 \in \mathbb{R}^d$ is the intercept and $\{ \epsilon_t\} \subset \mathbb{R}^d$ are i.i.d. errors. By taking expectation on both sides, we observe that $ E[z_t] - E[w_t] = \beta_0$. In some sense, this corresponds to a special case of  Model \eqref{model1} where  $p=1$ and the single  ``slope" is  $\alpha_1=1$ as then one obtains $y_t \ominus \mu_y = x_t \ominus \mu_x  + \varepsilon_t$ and its Euclidean counterpart  $ z_t - E[z_t] = w_t - E[w_t] +\epsilon_t$, which is equivalent to $ z_t - w_t = \beta_0 + \epsilon_t$.

As alternative to the SAR model (\ref{model1})  we also consider a second model that is based on the  spherical differences of consecutive observations.    This difference based  spherical autoregressive model (DSAR) is given by 
\begin{align} \label{model2}
R_{t} - \mu_R = \alpha_1  (R_{t-1}-\mu_R) + \cdots + \alpha_p (R_{t-p}-\mu_R) + \varepsilon_t, \text{ where } R_{t} = x_{t+1} \ominus x_{t},
\end{align}
where as before,  $\alpha_1, \cdots, \alpha_p \in \mathbb{R}$, $\mu_R = E[R_t]$ and  $\{ \varepsilon_t \} \subset \mathscr{S}(\mathcal{H})$ are i.i.d random innovations with mean 0. 

Differencing is an inherent feature of DSAR models and is  a common technique to reduce trend and seasonality for time series in Euclidean space. It  may also be useful for some  spherical time series.  For example, the US energy mix compositional time series, which we will discuss further in  Section 5.2, shows a  trend over the years, as more clean energy is generated each year and coal/petroleum fuels are increasingly phased out.  

Regarding the existence of stationary solutions of the proposed SAR model, the following result is a consequence of 
Theorem 3.3 of \cite{zhan:21}.\vs

\no {\bf Theorem 2.} {\it   
Assuming that $\{R_t:t \in \mathbb{N}\}$ is stationary, $E \langle \varepsilon_t, \varepsilon_t \rangle_{\mathcal{H}\otimes \mathcal{H}} < \infty $ and the roots of $ \phi(z) = 1 - \alpha_1 z - \cdots - \alpha_p z^p $ are outside the unit circle, then 
\begin{align*}
R_t - \mu_R = \sum_{i=0}^{\infty} \psi_i \varepsilon_{t-i}
\end{align*}
is a unique stationary solution of
\begin{align*}
R_t -\mu_R = \alpha_1 (R_{t-1} - \mu_R) + \cdots + \alpha_p ( R_{t-p} - \mu_R) + \varepsilon_t, \; t \in \mathbb{N},
\end{align*}
where $\{\psi_t\}$ is absolutely summable and determined by $1/\phi(z) = \sum_{i=0}^{\infty} \psi_i z^i$.
}\\

\bc {\bf \sf \Large 4.\quad Estimation and Prediction}\sm \ec \rs

\noindent { \sf 4.1 \quad Estimation}

\no We use Yule-Walker type estimators for the estimation of the coefficients  $\alpha_1, \alpha_2, \cdots, \alpha_p$ of the SAR and DSAR models.  Setting
$$ 
\lambda_{k} = E [ \langle R_1 - \mu_R, R_{k+1} - \mu_R  \rangle_{\mathcal{H} \otimes \mathcal{H}} ],
$$
it is straight-forward to check that the model parameters satisfy 
\begin{align*}
\left(
\begin{array}{c}
 \lambda_1  \\
 \lambda_2  \\
 \vdots \\
 \lambda_p
\end{array}
\right) = 
\left(
\begin{array}{cccc}
\lambda_0 & \lambda_1 & \cdots & \lambda_{p-1}  \\
\lambda_1 & \lambda_0 & \cdots & \lambda_{p-2} \\
\vdots & \vdots  &   & \vdots \\
\lambda_{p-1} & \lambda_{p-2} & \cdots & \lambda_{0} 
\end{array} \right)
\left(
\begin{array}{c}
 \alpha_1  \\
 \alpha_2  \\
 \vdots \\
 \alpha_p
\end{array}
\right).
\end{align*}
Replacing  $\lambda_k$ by  sample estimates 
\begin{align} \label{la-est} 
\widehat{\lambda}_k = \frac{1}{n-k} \sum_{t=1}^{n-k} \langle R_t - \widehat{\mu}_{R}, R_{t+k} - \widehat{\mu}_{R}  \rangle_{\mathcal{H} \otimes \mathcal{H}}, \quad  \widehat{\mu}_{R} = \frac{1}{n} \sum_{t=1}^n R_t
\end{align}
then suggests   the following estimates $\widehat{\alpha}_1, \cdots, \widehat{\alpha}_p$ for the model parameters   ${\alpha}_1, \cdots, {\alpha}_p,$
\begin{align}  \label{alph-hat} 
\left(
\begin{array}{c}
 \widehat{\alpha}_1  \\
 \widehat{\alpha}_2  \\
 \vdots \\
 \widehat{\alpha}_p
\end{array}
\right)
 = 
\left(
\begin{array}{cccc}
\widehat{\lambda}_0 & \widehat{\lambda}_1 & \cdots & \widehat{\lambda}_{p-1}  \\
\widehat{\lambda}_1 & \widehat{\lambda}_0 & \cdots & \widehat{\lambda}_{p-2} \\
\vdots & \vdots  &   & \vdots \\
\widehat{\lambda}_{p-1} & \widehat{\lambda}_{p-2} & \cdots & \widehat{\lambda}_{0} 
\end{array} \right)^{-1}
\left(
\begin{array}{c}
 \widehat{\lambda}_1  \\
 \widehat{\lambda}_2  \\
 \vdots \\
 \widehat{\lambda}_p
\end{array}
\right) .
\end{align}
Writing  $ \bm{\lambda} = (\lambda_0, \lambda_1, \cdots, \lambda_p)^T $ and $ \widehat{\bm{\lambda}} = (\widehat{\lambda}_0, \widehat{\lambda}_1, \cdots, \widehat{\lambda}_p)^T $, we next establish asymptotic normality for  $ \widehat{\bm{\lambda}} $. \vs

\no {\bf Theorem 3.} {\it   
Under the assumptions of Theorem 2, 
it holds that
\bea
    \sqrt{n} ( \widehat{\bm{\lambda}} - \bm{\lambda} ) \rightarrow^d  N(0, V),\quad\quad V = \left( \sum_{h = -\infty}^{\infty} \Gamma^h_{u,v} \right)_{u,v=0, 1, \cdots, p},
    \eea
where, setting
$\kappa(u) = \sum_{i= -\infty}^{\infty} \psi_{i} \psi_{i+u}$,
\begin{multline*}
    \Gamma^h_{u,v} =  \left( E[ \langle \varepsilon_{1} , \varepsilon_{1}  \rangle^2 ] - (E[ \langle \varepsilon_{1} , \varepsilon_{1}  \rangle ])^2- 2E[ \langle \varepsilon_{1} , \varepsilon_{2}  \rangle^2  ] \right) \sum_{i =-\infty}^{\infty} \psi_{i} \psi_{i+u} \psi_{i+h} \psi_{i+h+v}  \\
+  (E[ \langle \varepsilon_{1} , \varepsilon_{1}  \rangle ])^2 \kappa(u) \kappa(v) + E[ \langle \varepsilon_{1} , \varepsilon_{2}  \rangle^2  ] \left( \kappa(h)\kappa(h+v-u)   +  \kappa(h+v)\kappa(h-u) \right).
\end{multline*}
}
For the case where  $\{ \varepsilon_t \}$ are i.i.d random innovations in $\mathbb{R}$, the $\Gamma_{u,v}^h$ are identical to those in   Bartlett's formula. To show the convergence of $\widehat{\alpha}_1, \cdots, \widehat{\alpha}_p$, we set 
\begin{align*}
    \Lambda = \left(
\begin{array}{cccc}
\lambda_0 & \lambda_1 & \cdots & \lambda_{p-1}  \\
\lambda_1 & \lambda_0 & \cdots & \lambda_{p-2} \\
\vdots & \vdots  &   & \vdots \\
\lambda_{p-1} & \lambda_{p-2} & \cdots & \lambda_{0} 
\end{array} \right) \text{ and } \widehat{\Lambda} = \left(
\begin{array}{cccc}
\widehat{\lambda}_0 & \widehat{\lambda}_1 & \cdots & \widehat{\lambda}_{p-1}  \\
\widehat{\lambda}_1 & \widehat{\lambda}_0 & \cdots & \widehat{\lambda}_{p-2} \\
\vdots & \vdots  &   & \vdots \\
\widehat{\lambda}_{p-1} & \widehat{\lambda}_{p-2} & \cdots & \widehat{\lambda}_{0} 
\end{array} \right).
\end{align*}
Suppose that $\text{det}(\Lambda) \neq 0$, it then follows from the continuous mapping theorem that $\widehat{\Lambda}^{-1} \rightarrow^p \Lambda^{-1}$ and thus by Theorem 3, we have \vs

\no {\bf Corollary 1.} {\it   
Under the assumptions of Theorem 2, if $\emph{\text{det}}(\Lambda) \neq 0$,}
\begin{align*}
\sqrt{n} \left(
 \left(
\begin{array}{c}
 \widehat{\alpha}_1  \\
 \vdots \\
 \widehat{\alpha}_p
\end{array}
\right)-   \left(
\begin{array}{c}
 \alpha_1  \\
 \vdots \\
 \alpha_p
\end{array}
\right) \right) \rightarrow^d N\left(0, \Lambda \widetilde{V} (\Lambda^{T})^{-1} \right), \text{ where } \widetilde{V} = \left( \sum_{h = -\infty}^{\infty} \Gamma^h_{u,v} \right)_{u,v= 1, \cdots, p}.
\end{align*}

We note that in  applications involving distributional time series the distributions and specifically the density functions $f_t$ 
are usually not directly observed  and must be inferred from available  samples of size $N_t, \quad \{ z_{i,t} \in \mathbb{R}^D : i=1,2, \cdots, N_t \} \sim^{i.i.d} f_t$ that they generate. The random mechanism that generates the samples is assumed to be independent from the mechanism that generates the random distributions. 

To assess  the impact of this preliminary estimation step, we provide one requires additional assumptions as follows: All densities  $f_t$ have the same compact domain $A \subset \mathbb{R}^D$ and are continuously differentiable on their support;  there is a sequence $N \rightarrow \infty$ such that  $N_t \ge N$ for all $t$; 
 there exists a constant $M$ such that for all $t$, $ \sup_{a \in A} |f_t(a)| $, $ \sup_{a \in A} 1/|f_t(a)| $, $ \sup_{a \in A} \|f_t'(a)\| $ are all bounded by $M$, where  $\|f_t'(a)\|$ is the norm of the gradient vector. Extending  the arguments and construction in  \cite{mull:16:1} to the case of multivariate distributions leads to density estimators $\hat{f}_t$ that satisfy
\begin{align*}
    \sup_t P \left(  \sup_{a \in A} \left|\hat{f}_t(a) -f_t(a)\right| > c_1 N^{-c_2} \right) \rightarrow 0
\end{align*}
for constants $c_1, c_2>0$, where $c_2$ depends on the dimension of the distributions and decreases when the dimension increases. One can then show that substituting $\hat{f}_t$ for $f_t$ in $R_t$ in models (\ref{model1}) and (\ref{model2}) and choosing the sample size $N=N(n)$ available for the estimation of each density $f_t$ such that $N^{-c}=o_p(n^{-1/2})$ implies that Theorem 3 and Corollary 1 still hold when using estimated instead of true densities in the fitting of the SAR models under these additional assumptions.\vs

\noindent { \sf 4.2 \quad Prediction}

With  estimates $\widehat{\alpha}_1, \cdots, \widehat{\alpha}_p$ based on data sequence $\{R_1, \cdots, R_n\}$ in hand, the  prediction for the skew-symmetric operator at time $n+1$ is 
\begin{align*}
    \widehat{R}_{n+1} =\widehat{\mu}_R + \widehat{\alpha}_1 (R_{n} - \widehat{\mu}_R) + \cdots + \widehat{\alpha}_p (R_{n-p+1} - \widehat{\mu}_R),
\end{align*}
with a slight abuse of notation, as  in  model DSAR,  the sequence of observations available for the prediction is of length $n+1$, i.e.,  $\{x_{i}:i=1,2, \cdots, n+1\}$, whereas in model  SAR  it is of length $n$.  Once $\widehat{R}_{n+1}$ has been obtained, the prediction of the next observation in the original time series is $\widehat{x}_{n+1}:= \exp(\widehat{R}_{n+1}) \mu_x$ when modeling with SAR and  $\widehat{x}_{n+2}:= \exp(\widehat{R}_{n+1}) x_{n+1}$ for DSAR.

For a distributional time series of $D$-dimensional distributions (or density functions), we set $\mathcal{H} = \{ f:\mathbb{R}^D  \rightarrow \mathbb{R}\;|\; \int_{\mathbb{R}^D} f^2(a) da < \infty \}$ with inner product $\langle f,g \rangle_{\mathcal{H}} = \int_{\mathbb{R}^D} f(a)g(a) da$ and  require the predictions to be constrained in the positive orthant $\mathcal{H}_{+}:= \{ f \in \mathcal{H}: f(a) \geq 0 \text{ for all }a \in\mathbb{R}^D \}$. Similarly, for compositional time series, $\mathcal{H} = \mathbb{R}^{d}$ and the prediction is constrained to lie in  $\mathcal{H}_{+} = \{ \mathbf{z} = (z_1, \cdots, z_d)^T \in \mathbb{R}^d : z_i \geq 0 \text{ for all }i \}$. Writing  $x_{\text{rot}} = \exp(Q) x$  for the rotation  $\exp(Q) $ of  $x \in \mathcal{H}_{+}$,   we use projection operators to enforce the constraint; see \cite{mull:21:4} and \cite{pego:21} 
for related projections in Wasserstein space. 

\begin{figure}
    \centering
    \begin{tikzpicture}
\draw[white] (-7,-3.5) -- (7, 3.5);

\draw[black, ultra thick] (-7, 0) -- (-1, 0);
\draw [gray!50] plot [smooth] coordinates {(-1,0) (-1.5,-3) (-6.5,-3) (-7,0)};
\node[circle, fill,inner sep=1.5,label=above left: $\text{Proj}^1(x_{\text{rot}})$] at (-4.4, 0)   (a) {};
\draw[black, ->, dashed, thick] (-5, -3) node[circle, fill,inner sep=1.5,label=left:$x$] (x) {} -- (-4, 2) node[circle, fill,inner sep=1.5,label=right:$x_{\text{rot}}$] (xrot) {};

\draw[black, ultra thick] (0, 0) -- (7, 0);
\draw[black, ->, dashed, thick] (1, -0.5) node[circle, fill,inner sep=1.5,label=left:$x$] (y) {} -- (6, 1) node[circle, fill,inner sep=1.5,label=right:$x_{\text{rot}}$] (yrot) {};

\draw [gray!50] plot [smooth] coordinates {(0,0) (0.5,-3) (6.5,-3) (7,0)};
\node[label={[gray!50] $\mathcal{H}_{+}$}] at (-2,-2.5)   (a) {};
\draw[black, ->, dashed, thick] (6,1) -- (6,0) node[circle, fill,inner sep=1.5,label=below: $\text{Proj}^2 (x_{\text{rot}})$]  (b) {};
\node[label={[gray!50] $\mathcal{H}_{+}$}] at (6,-2.5)   (a) {};
\end{tikzpicture}
    \caption{Illustration of the two projection operators Proj$^1$ and Proj$^2$.}
    \label{fig:projections}
\end{figure}
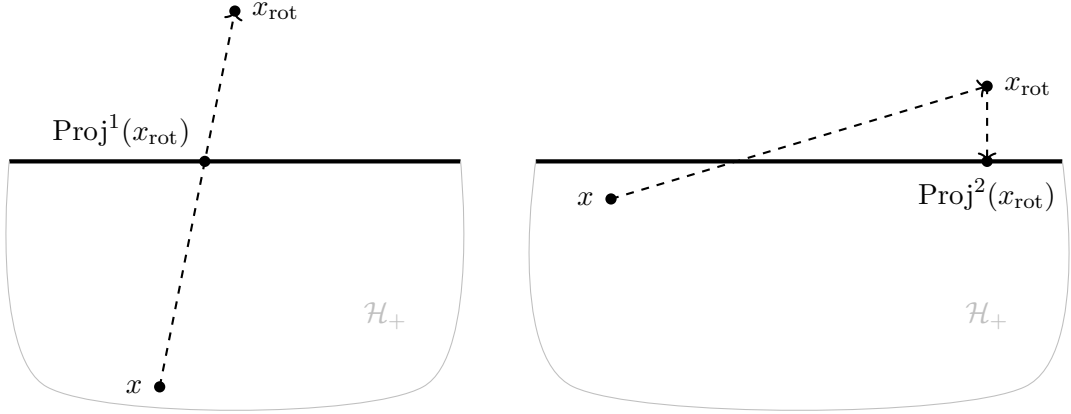

A first option is to use a projection operator  $\text{Proj}^1$ to rotate $x_{rot}$ back to the boundary of $\mathcal{H}_{+}$, i.e., $\text{Proj}^1(x_{\text{rot}}):= \exp(c_1 Q) x$, where $ c_1 = \sup\{ c: c \in [0,1] \text{ and } \exp(c Q) x \in \mathcal{H}_{+}\}$. A second option is the operator $\text{Proj}^2$ to  project $x_{\text{rot}}$ to the nearest point in $\mathcal{H}_{+}$, i.e., $\text{Proj}^2(x_{\text{rot}}) = \argmin_{y \in \mathcal{H}_{+}} \langle x_{\text{rot}}-y, x_{\text{rot}}-y \rangle$; see Figure \ref{fig:projections} for a schematic illustration.
We note that $\text{Proj}^1$ may be more useful for SAR, as all the predictions are based on rotations from the Fr\'{e}chet mean, which may be more likely to stay away from  the boundary of $\mathcal{H}_{+}$ under stationarity assumptions. Applying $\text{Proj}^1$ when constructing  predictions of SAR leads to  constrained  predictions that are closer to the Fr\'{e}chet mean than those obtained with $\text{Proj}^2$.  On the other hand, $\text{Proj}^1$ may be less useful for  DSAR, as 
one may obtain  $\widehat{x}_{n+2}:= \exp(\widehat{R}_{n+1}) x_{n+1} \approx x_{n+1}$. 
Therefore  $\text{Proj}^2$ appears to be  more suitable for DSAR. In the following, we use $\text{Proj}^1$ for SAR and $\text{Proj}^2$ for DSAR.\\

\bc {\bf \sf \Large 5.\quad Applications}\sm \ec \rs

\noindent { \sf 5.1 \quad Temperature data}

\no Global warming is expected to lead to  more heat waves in the summer. It  is then of interest to study and  model the time series of the bivariate distributions of daily minimum and maximum temperature. Extreme temperatures are associated with increased health and economic risks. The analysis reported here was inspired by \cite{bhat:21}. The temperature data we used have been recorded at airport weather stations in the U.S. over the years and are available at \url{https://www.ncdc.noaa.gov/cdo-web/search?datasetid=GHCND}. 

On the $i$th day of year $t=1990, \cdots, 2019$, we observe two temperatures $(z_{t, i}, w_{t, i})$, where $z_{t,i}$, $w_{t,i}$ are minimum and  maximum that temperature of each 24 hour period,  respectively. We assume that the distribution of 
$(z_{t, i}, w_{t, i})$ over the summer months in year $t$ has 
a  density  $f_t$ such that
\begin{align} \label{sample} 
    \{ (z_{t,i}, w_{t,i}): i =1,2, \cdots, N \} \sim^{i.i.d} f_{t},
\end{align}
where $N=122$ as we define the summer days to be June 1 to September 30. 

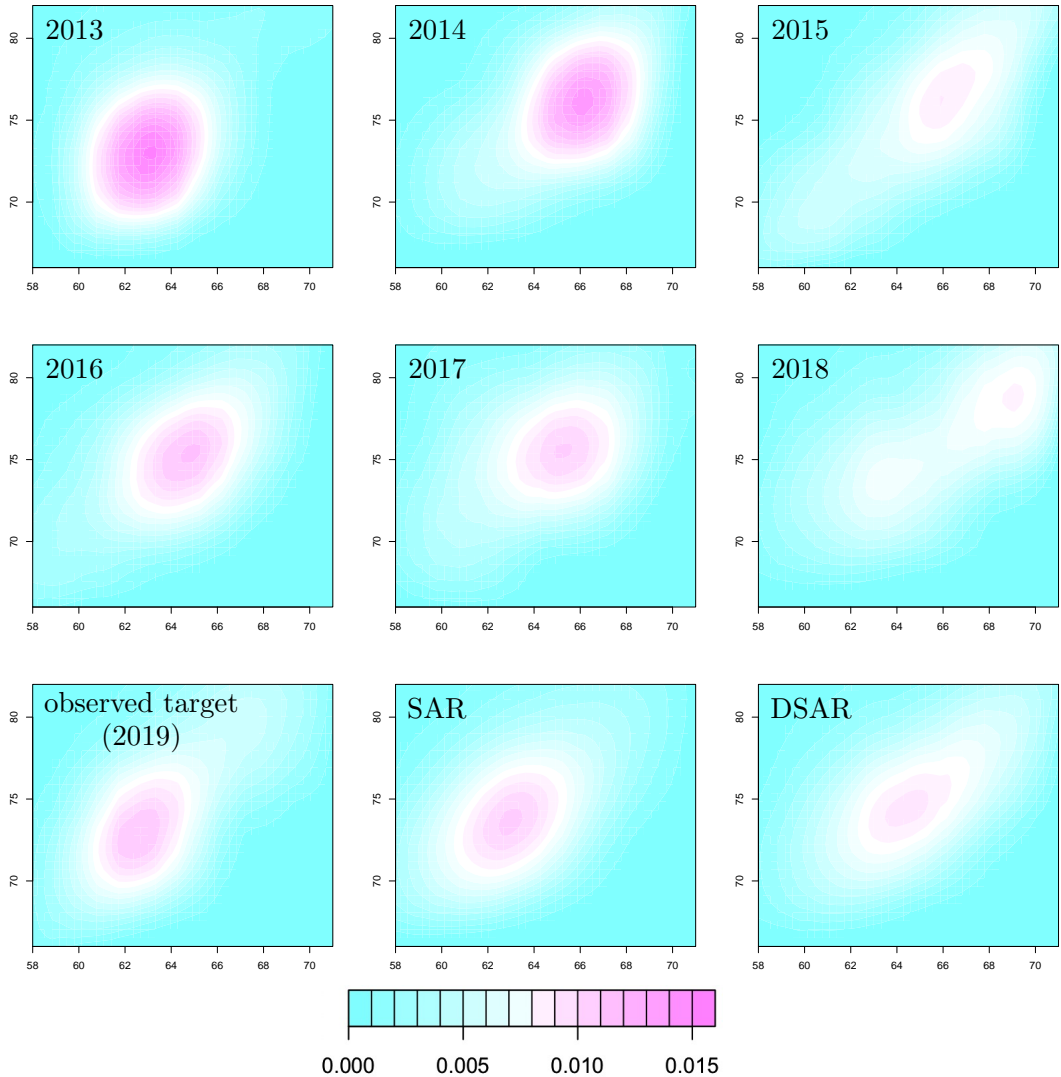
\begin{figure}
    \centering
    \begin{tikzpicture}
\matrix (m) [row sep = -2em, column sep = - 1.2em]{    
	 \node (p11) {\includegraphics[scale=0.33]{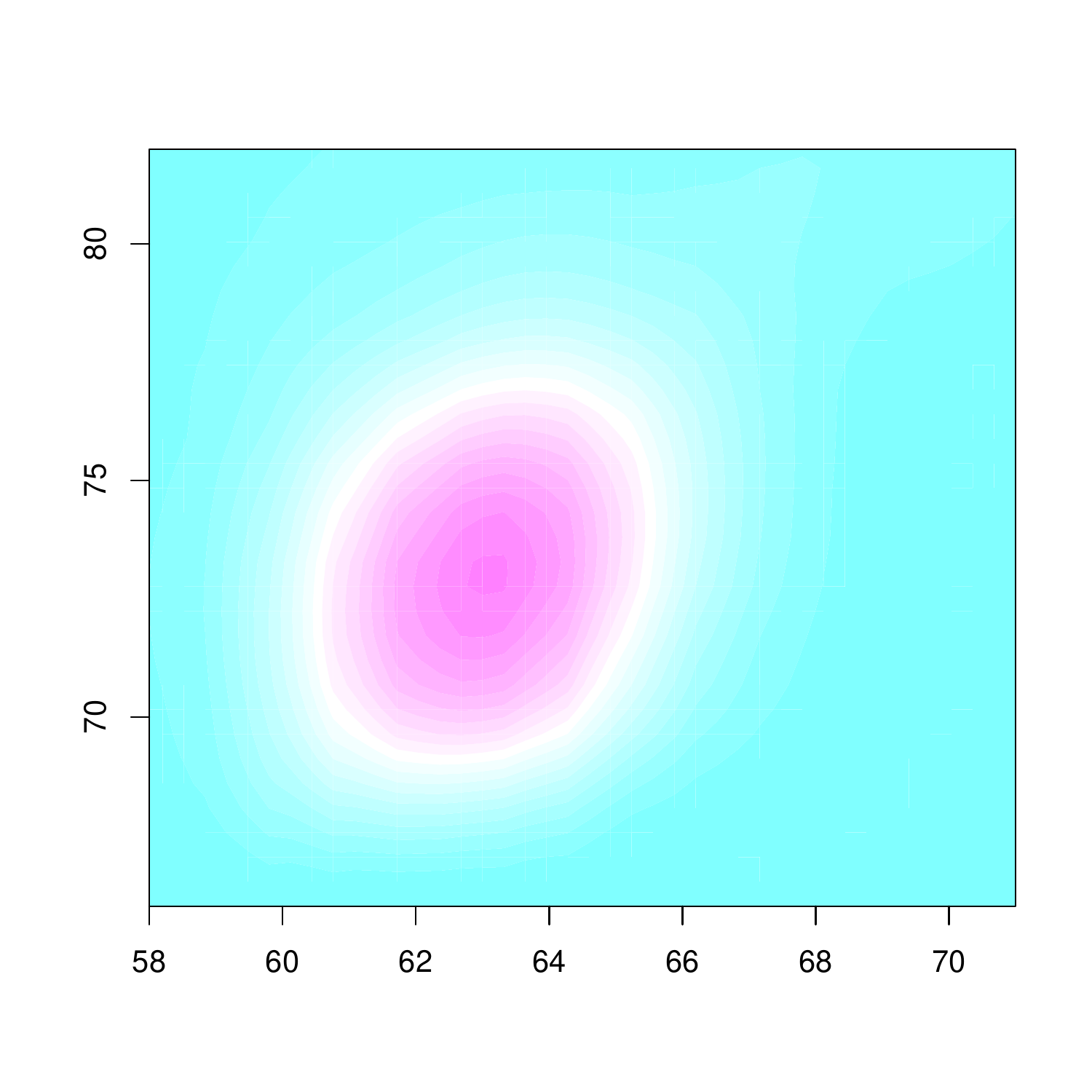}}; 
	 \node[above left = -1.4cm and -1.9cm of p11] (t12) {2013}; &
	 \node (p12) {\includegraphics[scale=0.33]{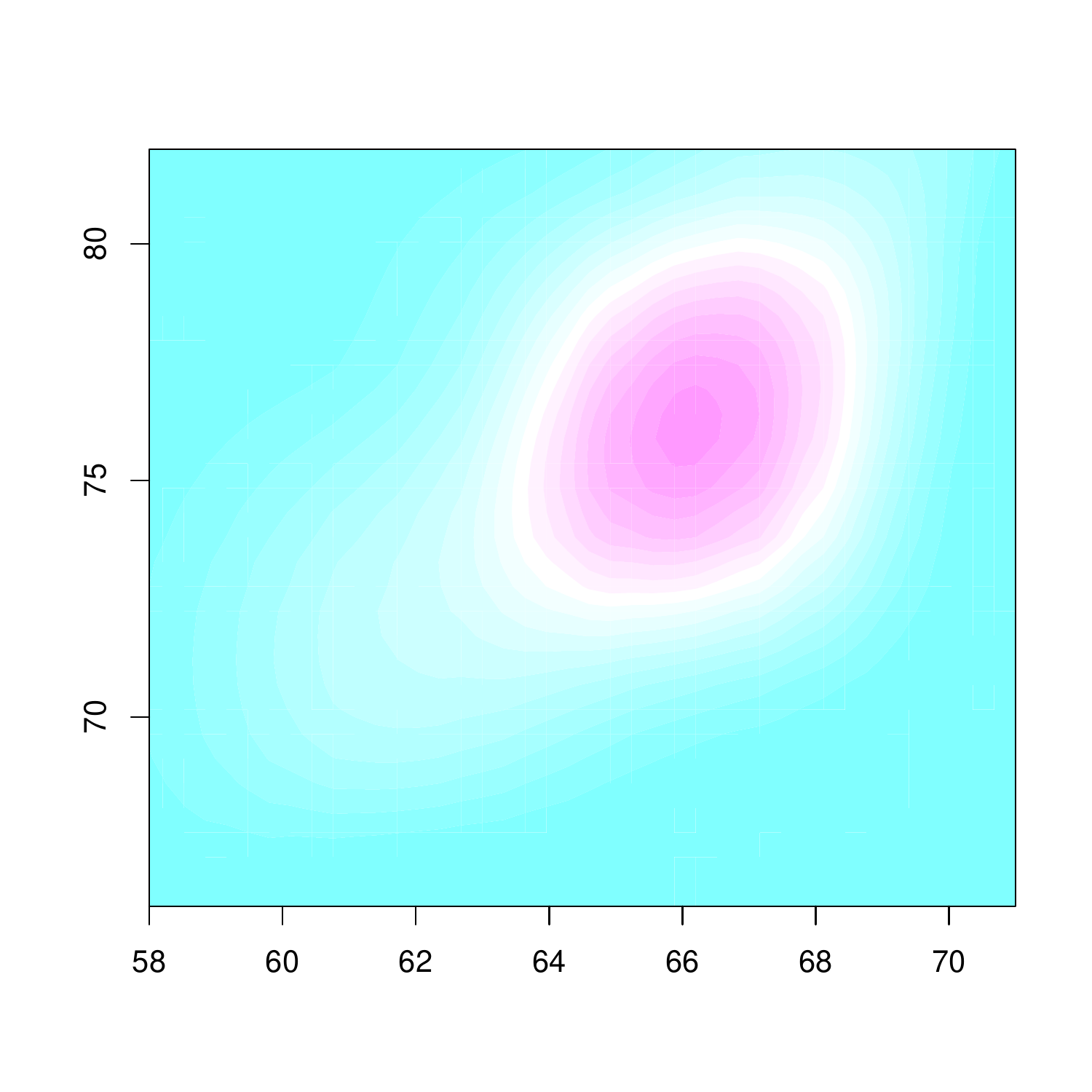}};
	 \node[above left = -1.4cm and -1.9cm of p12] (t21) {2014}; &
	 \node (p13) {\includegraphics[scale=0.33]{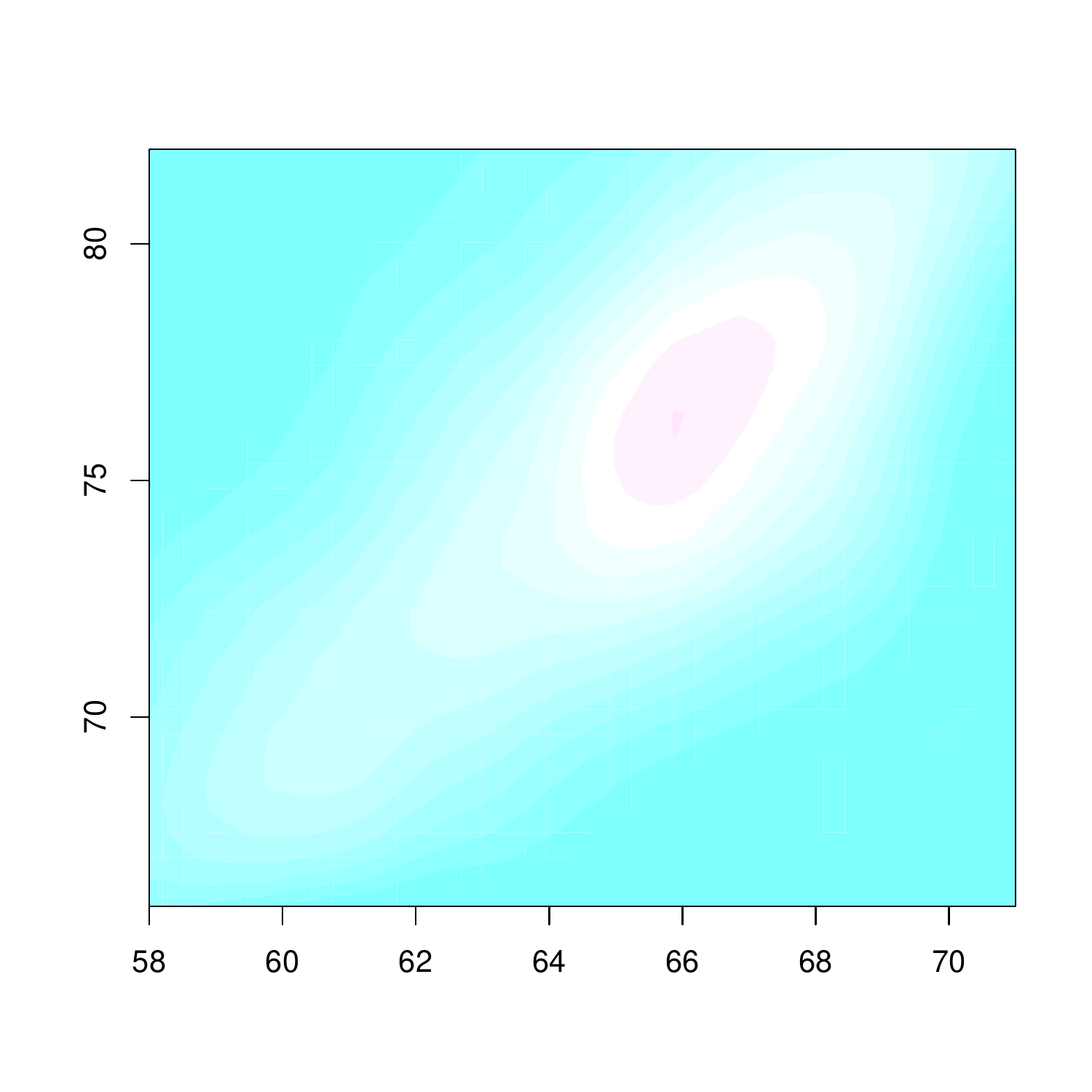}}; 
	 \node[above left = -1.4cm and -1.9cm of p13] (t22) {2015};
	 \\ 
	 \node (p21) {\includegraphics[scale=0.33]{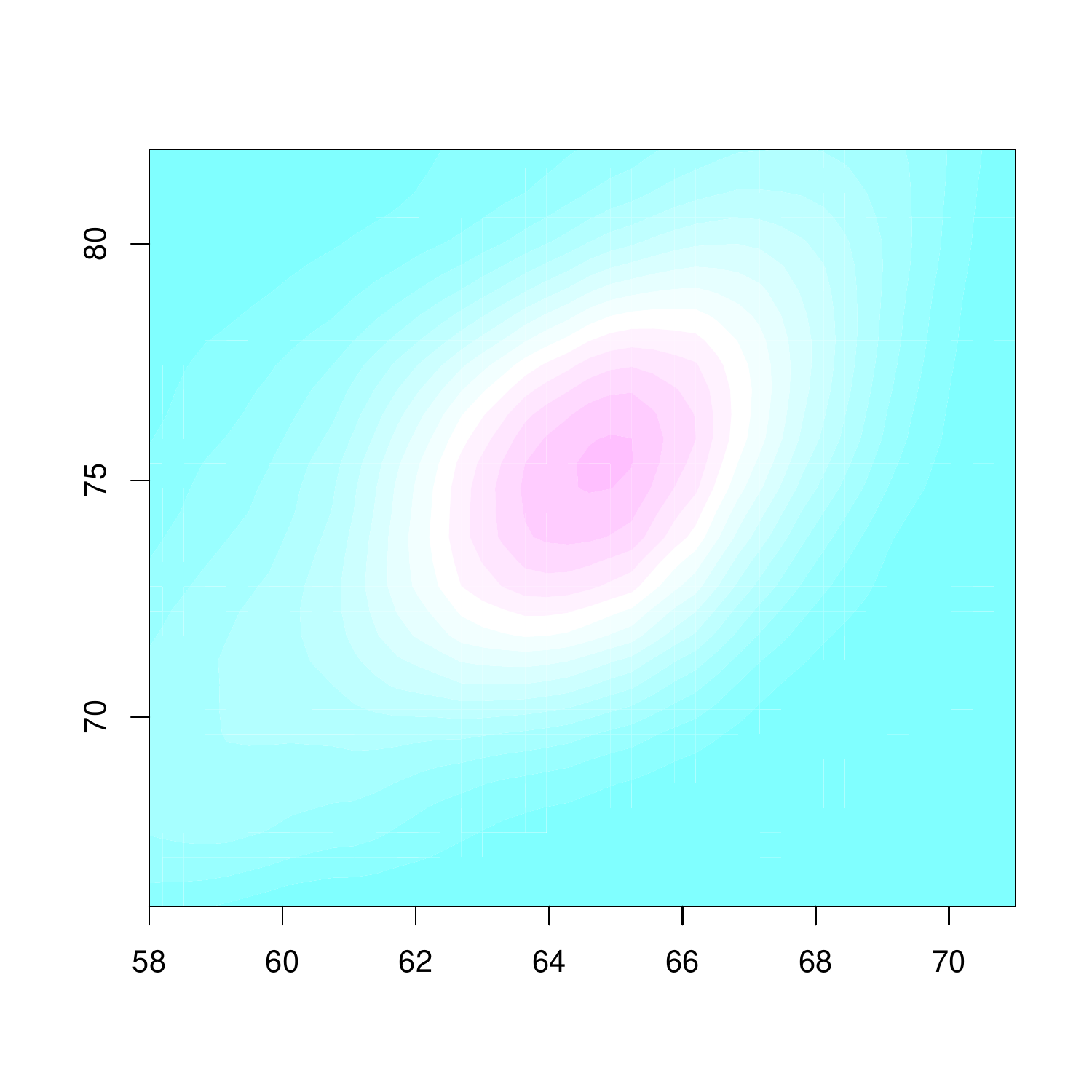}};
	 \node[above left = -1.4cm and -1.9cm of p21] (t31) {2016}; &
	 \node (p22) {\includegraphics[scale=0.33]{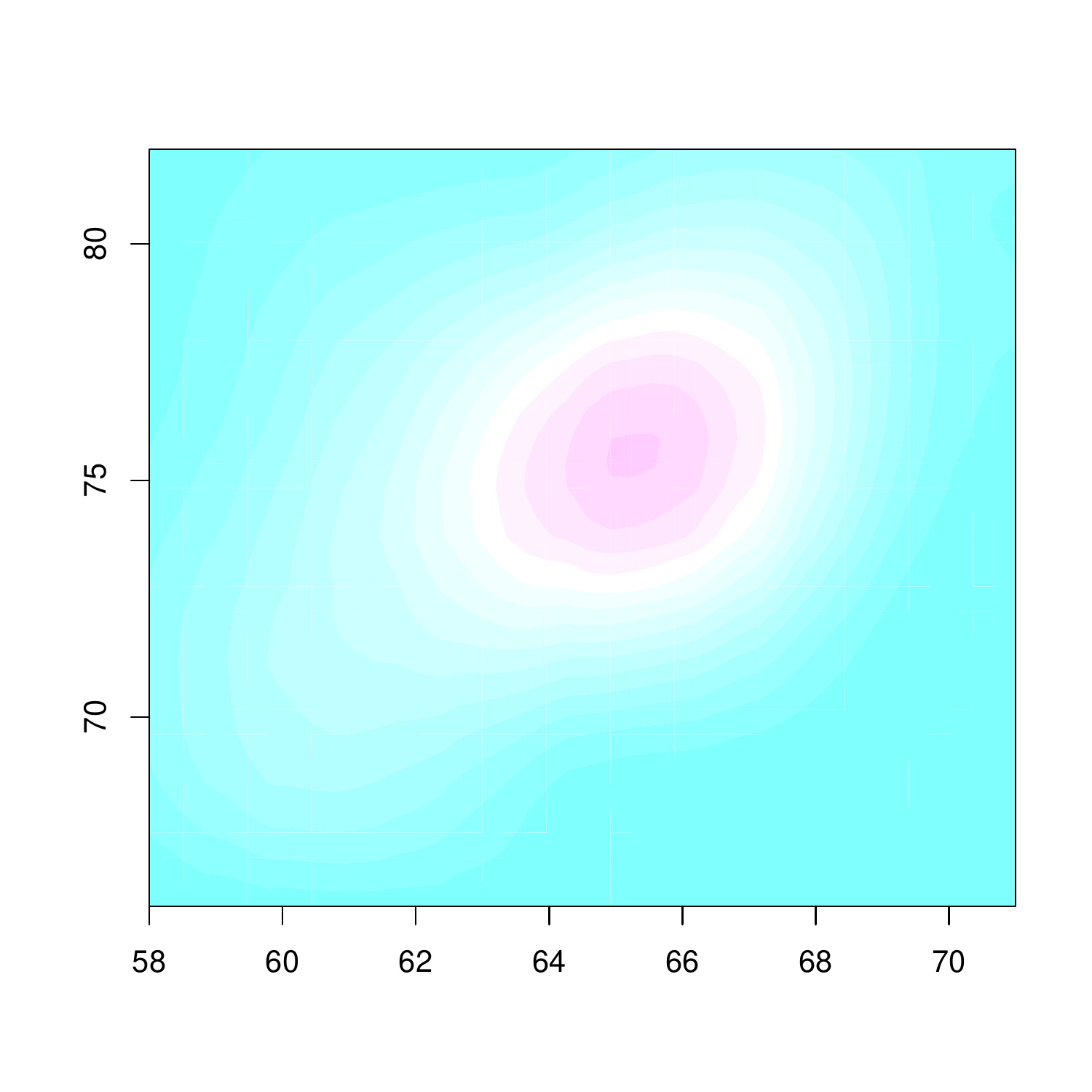}}; 
	 \node[above left = -1.4cm and -1.9cm of p22] (t32) {2017}; & 
	 \node (p23) {\includegraphics[scale=0.33]{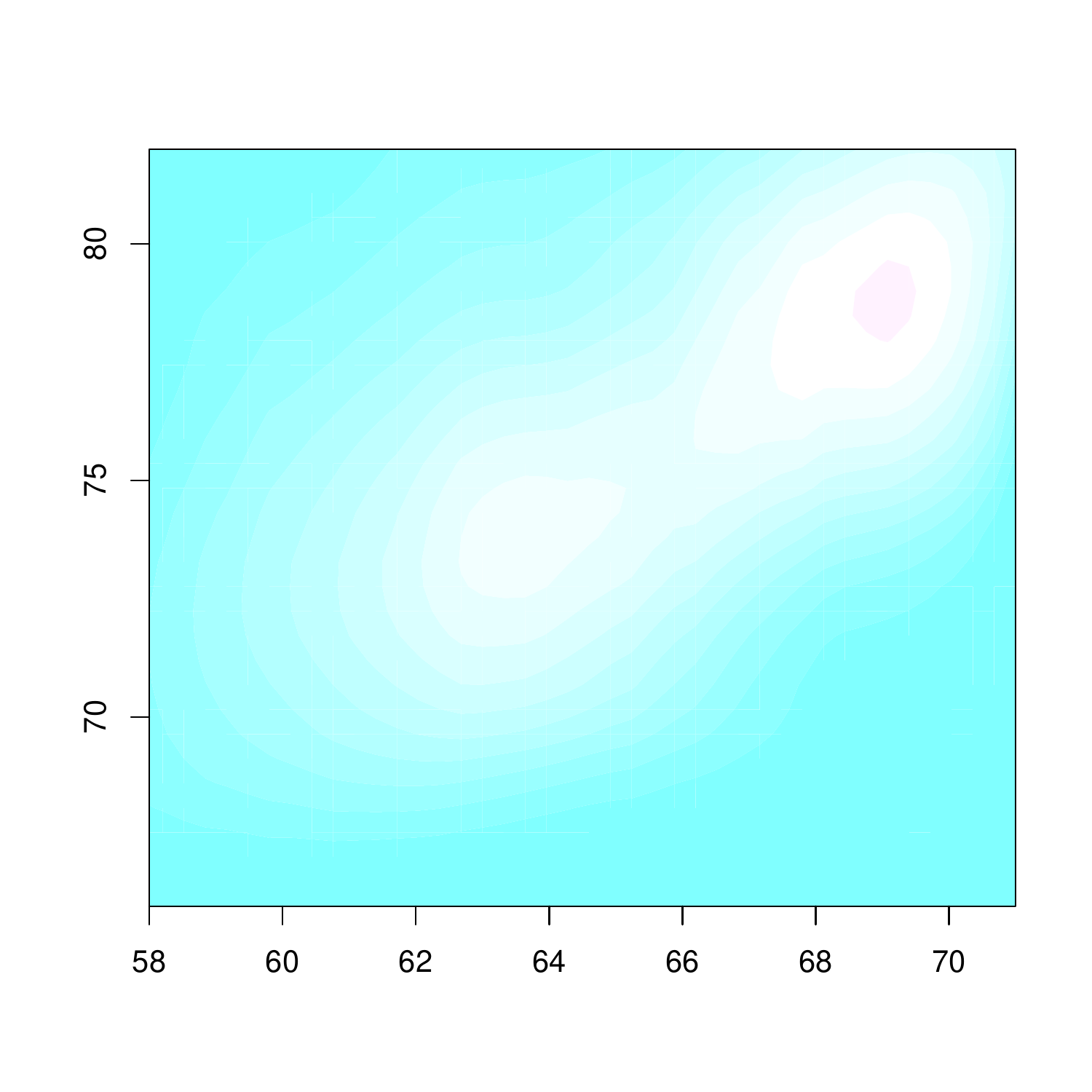}};
	 \node[above left = -1.4cm and -1.9cm of p23] (t11) {2018}; 
	 \\
	 \node (p31) {\includegraphics[scale=0.33]{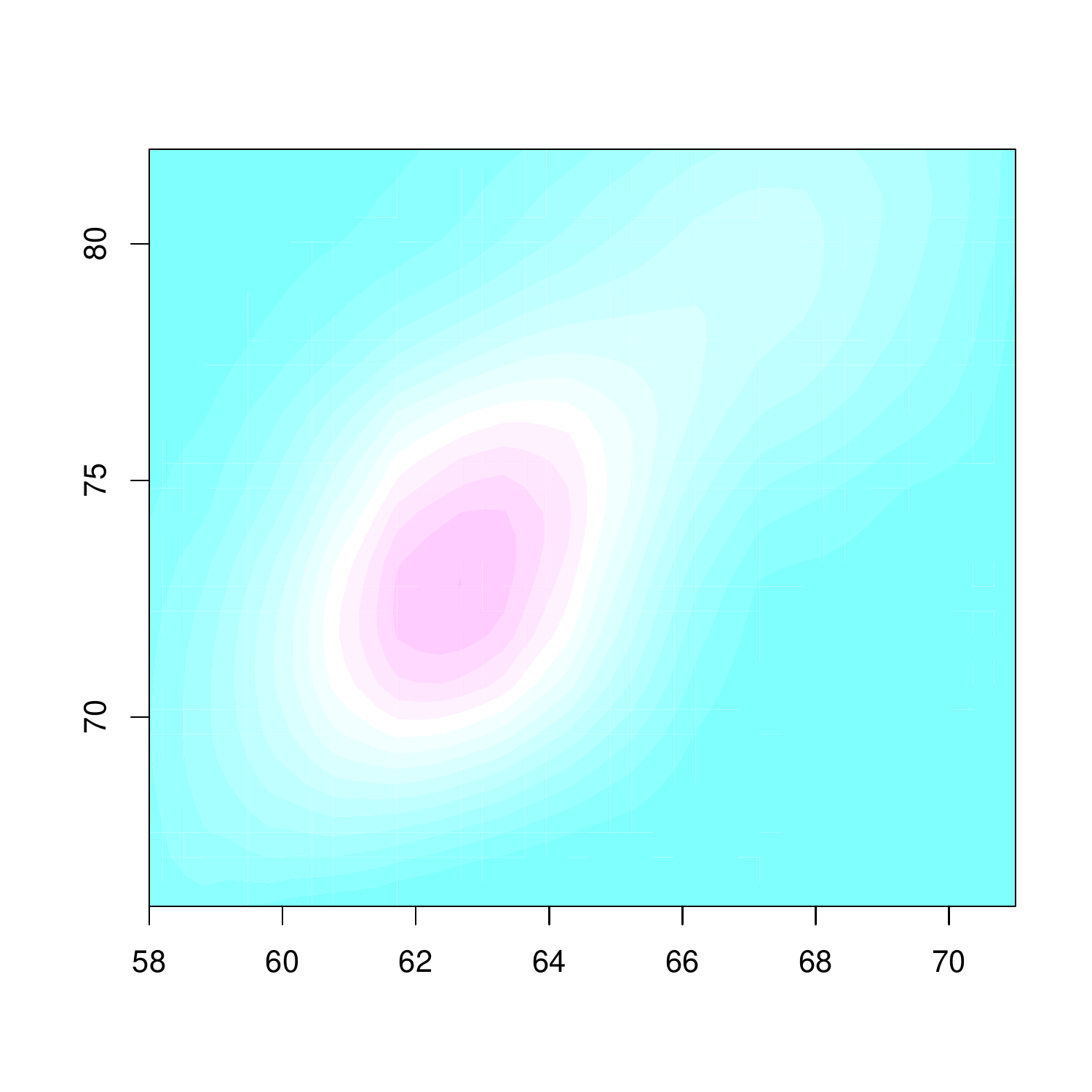}};
	 \node[above left = -1.4cm and -3.7cm of p22] (t22) {observed target};
	 \node[below = -0.2cm and 0cm of t22] (tt22) {(2019)};
	 & \node (p32) {\includegraphics[scale=0.33]{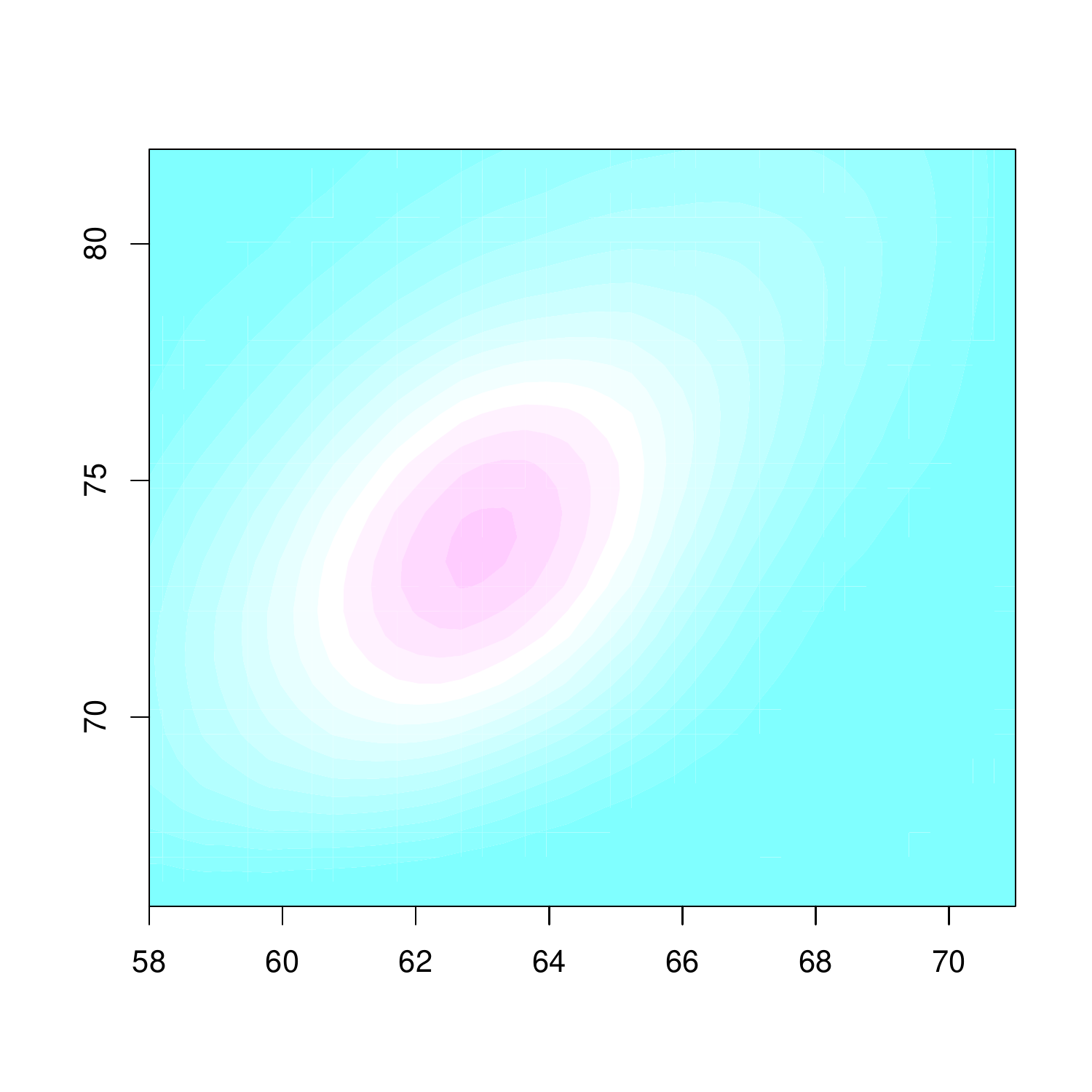}};
	 \node[above left = -1.4cm and -1.9cm of p32] (t31) {SAR};
	 & \node (p33) {\includegraphics[scale=0.33]{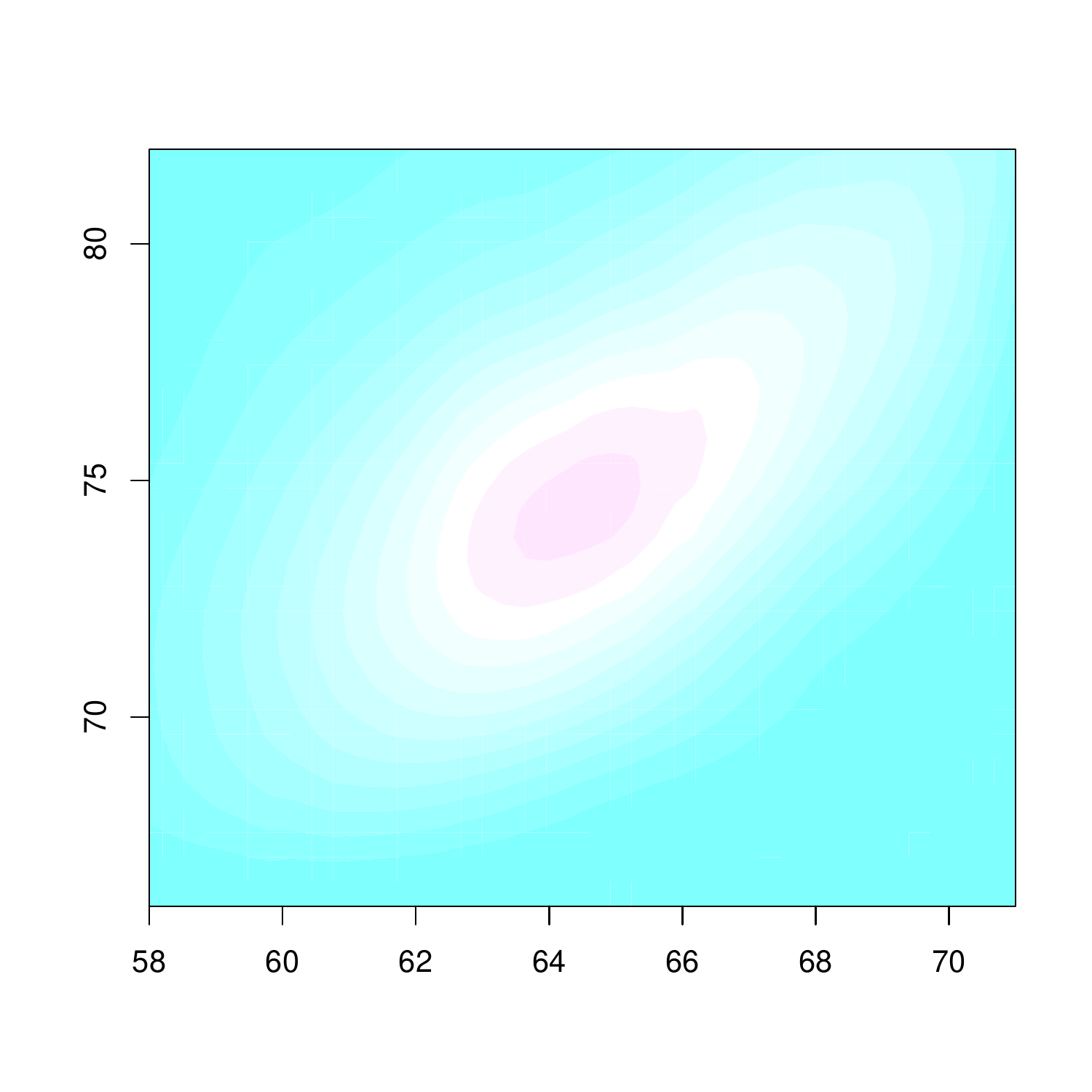}}; 
	 \node[above left = -1.4cm and -2.2cm of p33] (t32) {DSAR};
	 \\
};

\node[rotate=270,  below right=  -0.6cm and 0cm of p32] (legend) {\includegraphics[scale=0.25]{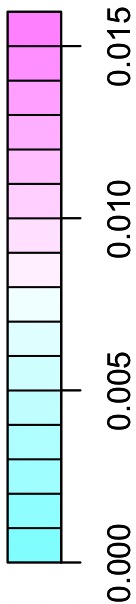}} ;
\node[above= -0.8cm and 0cm  of m] (title) {Los Angeles International Airport} ;
\end{tikzpicture}
    \caption{Contour plots of observed and predicted two-dimensional density functions for the distributional time series of temperatures as recorded at LAX. The top six panels show  the observed density functions in the training set. The bottom left panels show the observed distribution for  2019 (left);  the predicted density using SAR (middle), with Fisher-Rao distance between predicted and observed of 0.197; and  the predicted density using DSAR, with Fisher-Rao distance 0.236.}
    \label{fig:lax}
\end{figure}

\begin{figure}
    \centering
    \begin{tikzpicture}
\matrix (m) [row sep = -2em, column sep = - 1.2em]{    
	 \node (p11) {\includegraphics[scale=0.33]{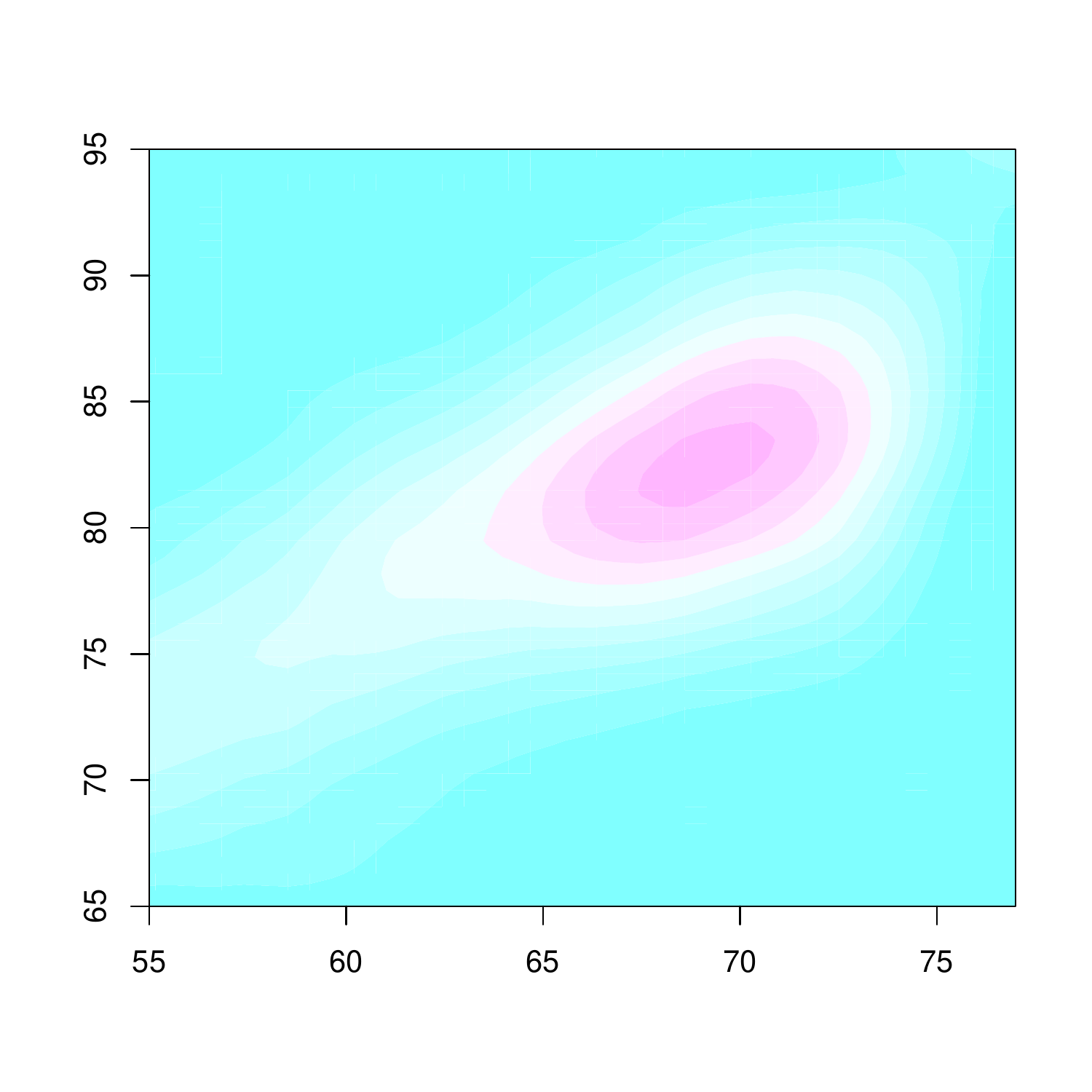}}; 
	 \node[above left = -1.4cm and -1.9cm of p11] (t12) {2013}; &
	 \node (p12) {\includegraphics[scale=0.33]{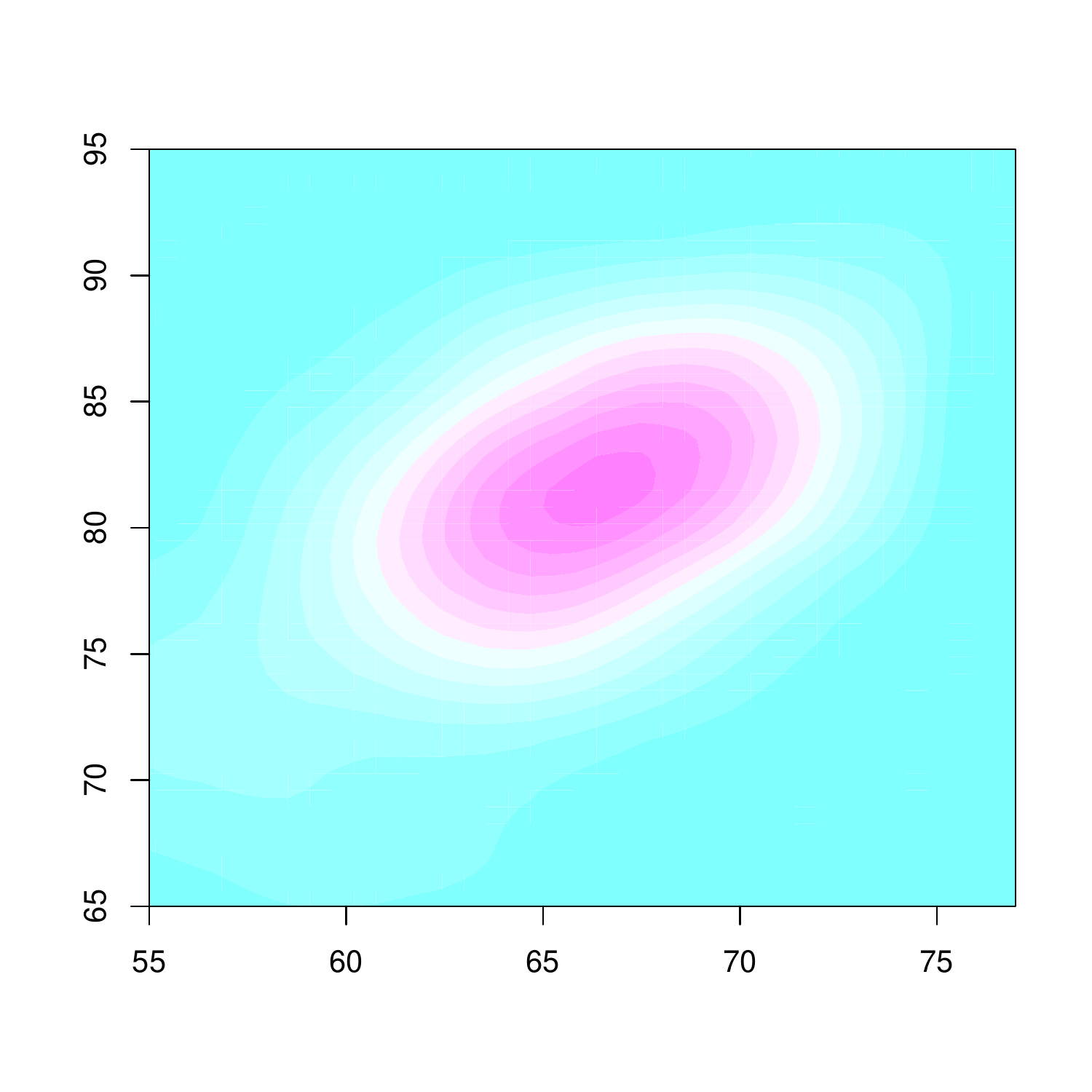}};
	 \node[above left = -1.4cm and -1.9cm of p12] (t21) {2014}; &
	 \node (p13) {\includegraphics[scale=0.33]{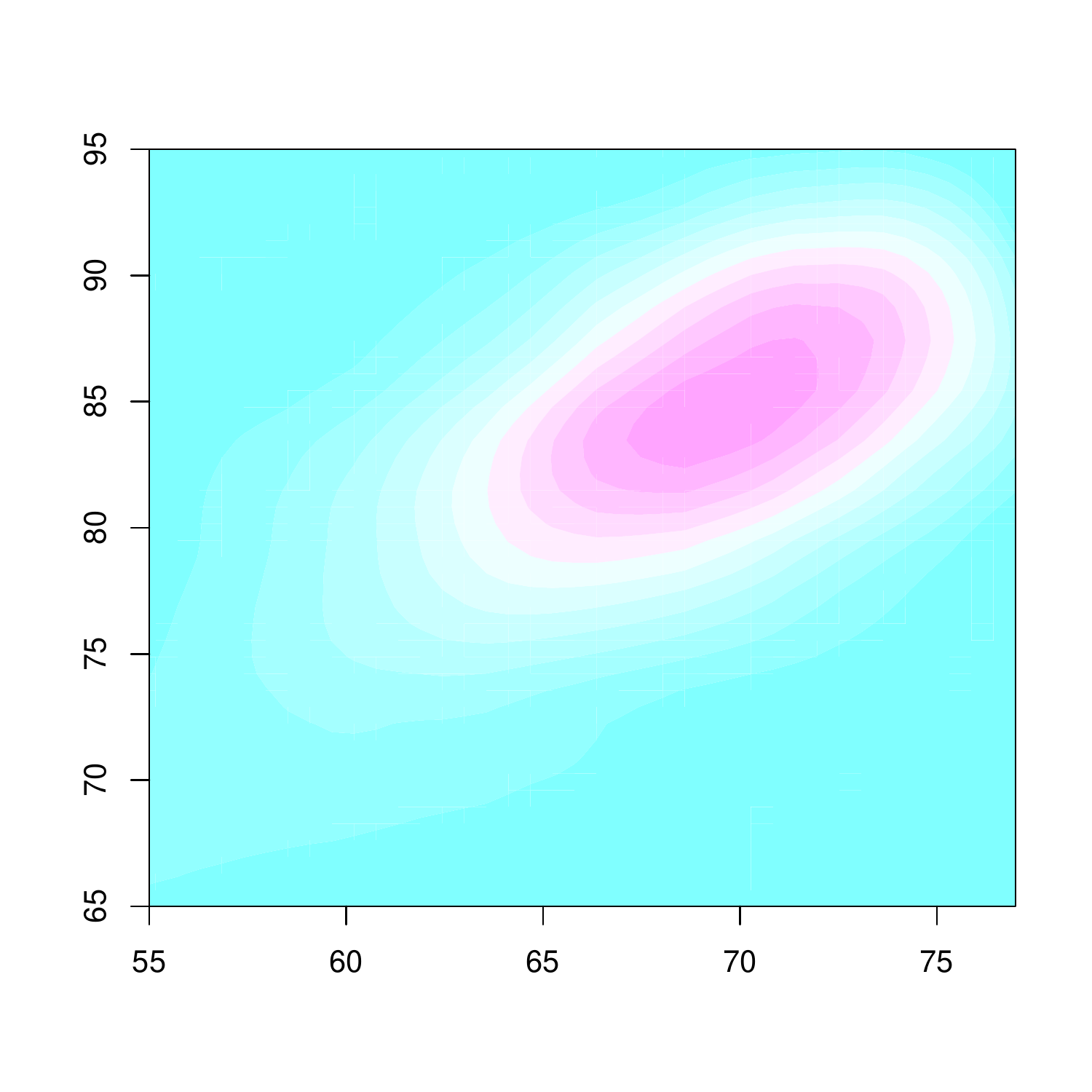}}; 
	 \node[above left = -1.4cm and -1.9cm of p13] (t22) {2015};
	 \\ 
	 \node (p21) {\includegraphics[scale=0.33]{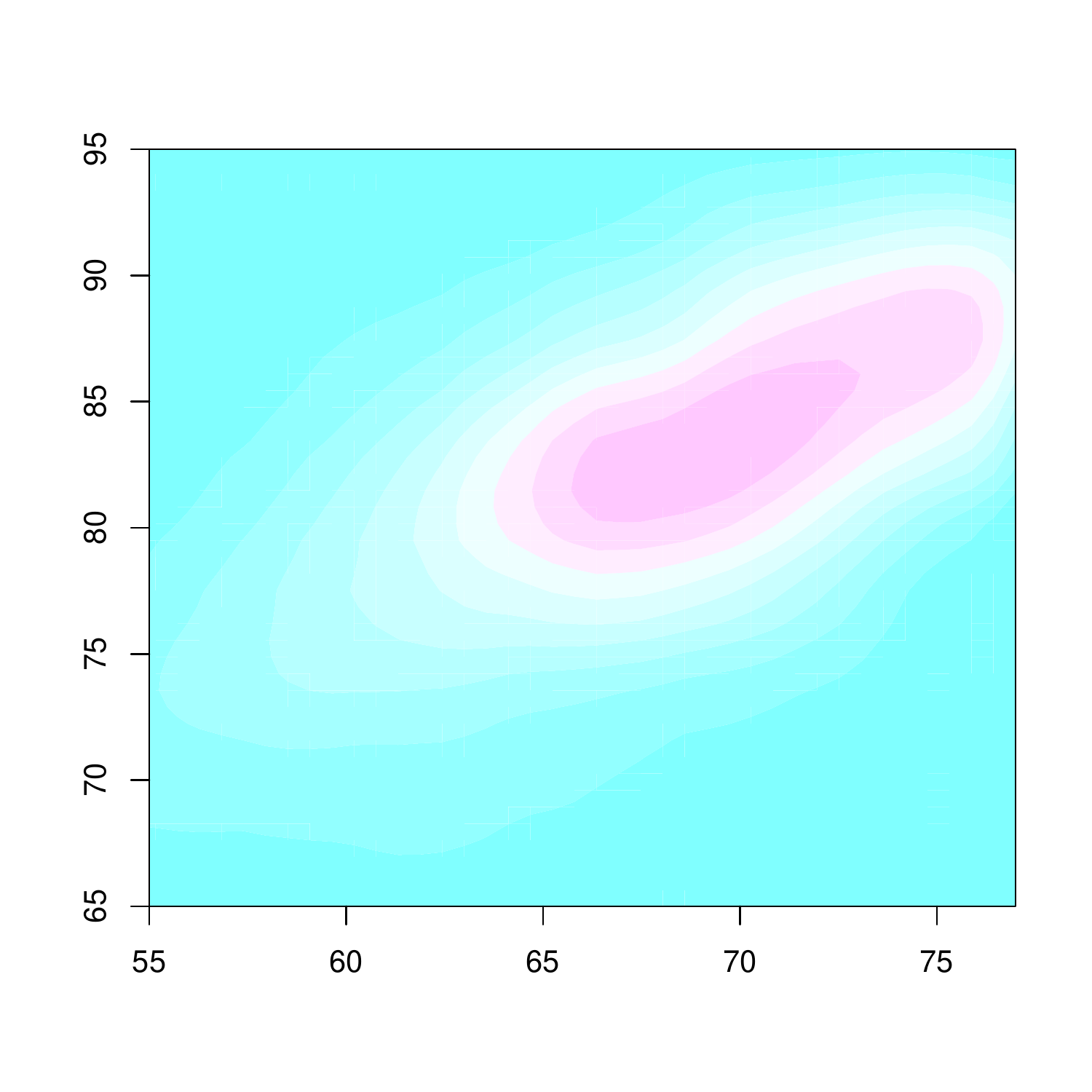}};
	 \node[above left = -1.4cm and -1.9cm of p21] (t31) {2016}; &
	 \node (p22) {\includegraphics[scale=0.33]{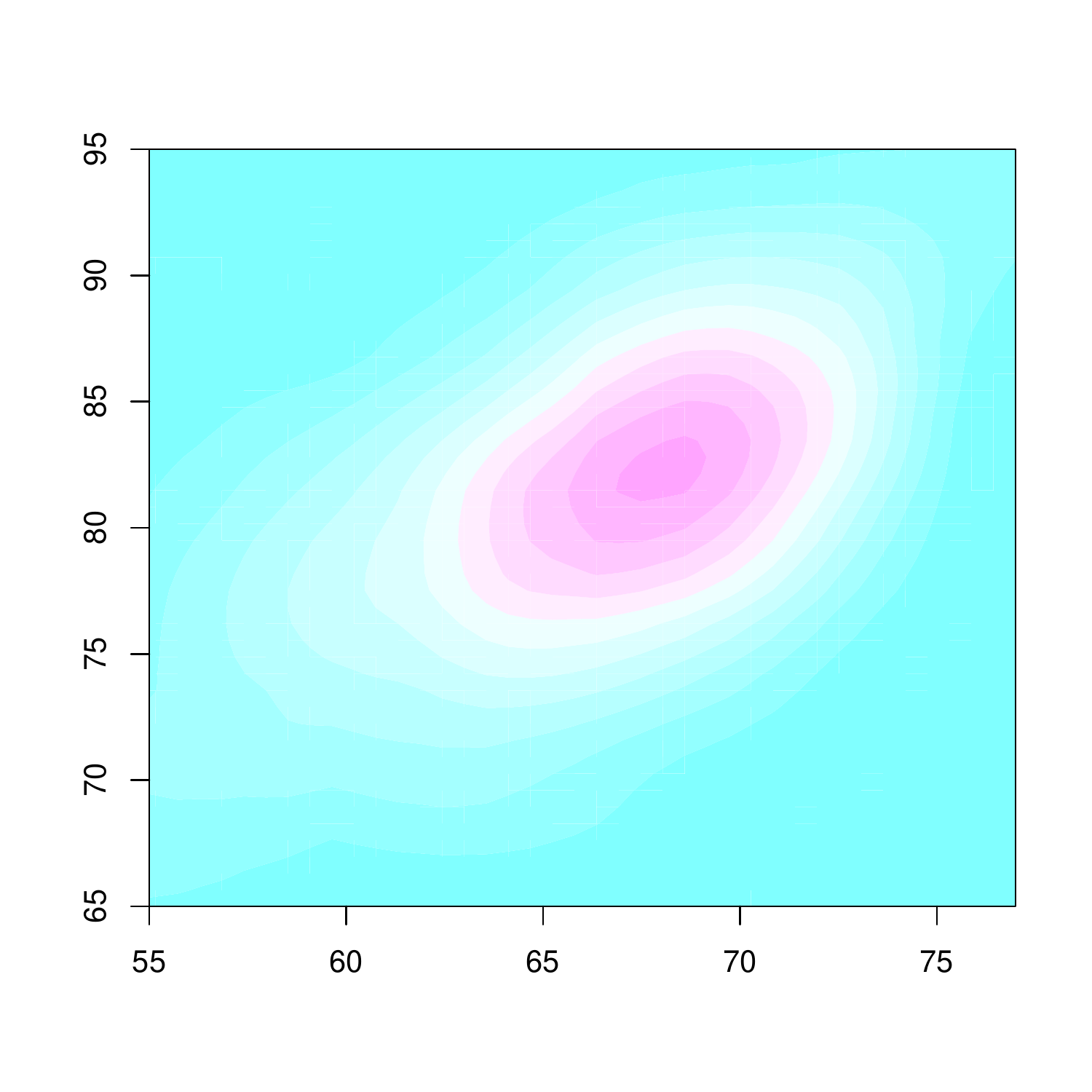}}; 
	 \node[above left = -1.4cm and -1.9cm of p22] (t32) {2017}; & 
	 \node (p23) {\includegraphics[scale=0.33]{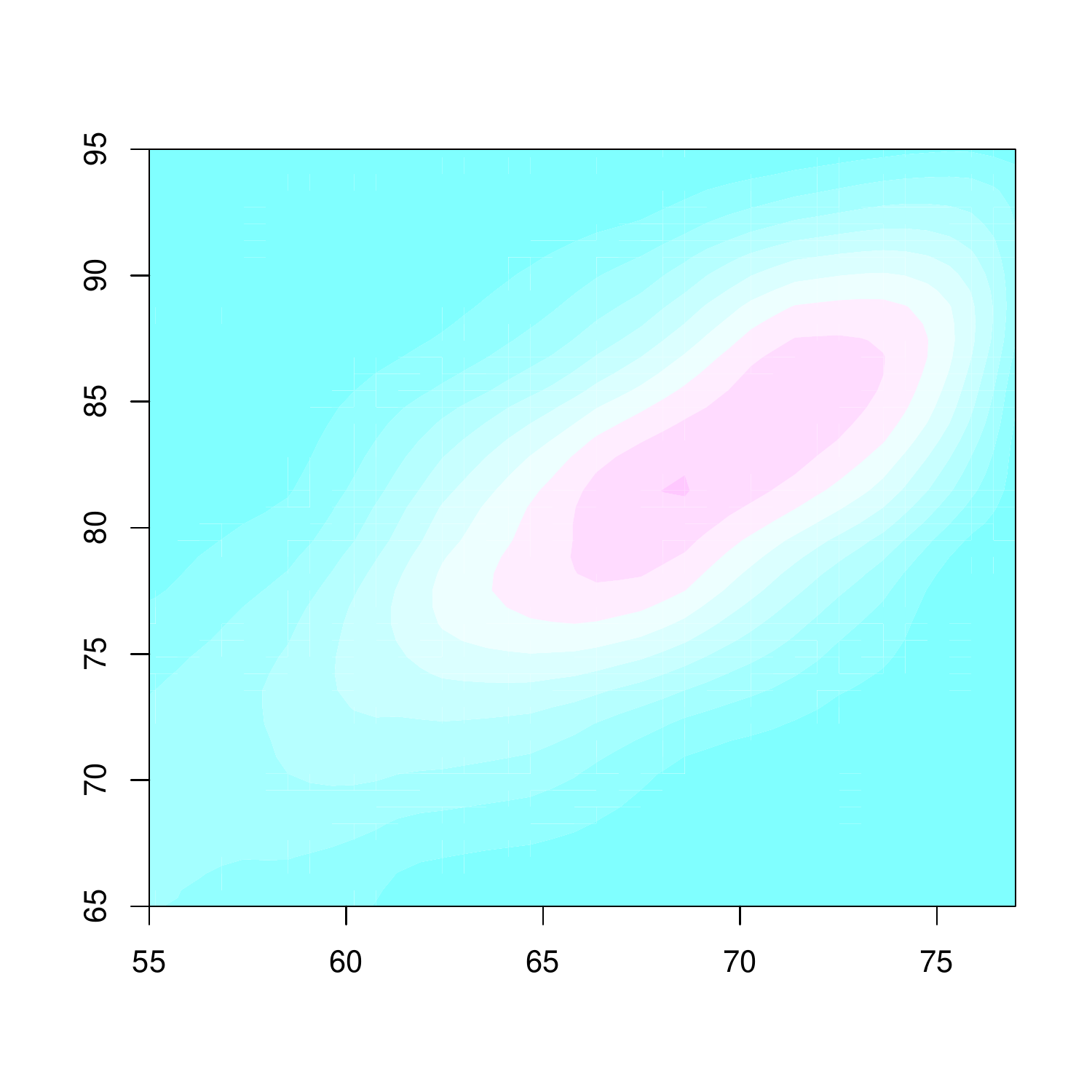}};
	 \node[above left = -1.4cm and -1.9cm of p23] (t11) {2018}; 
	 \\
	 \node (p31) {\includegraphics[scale=0.33]{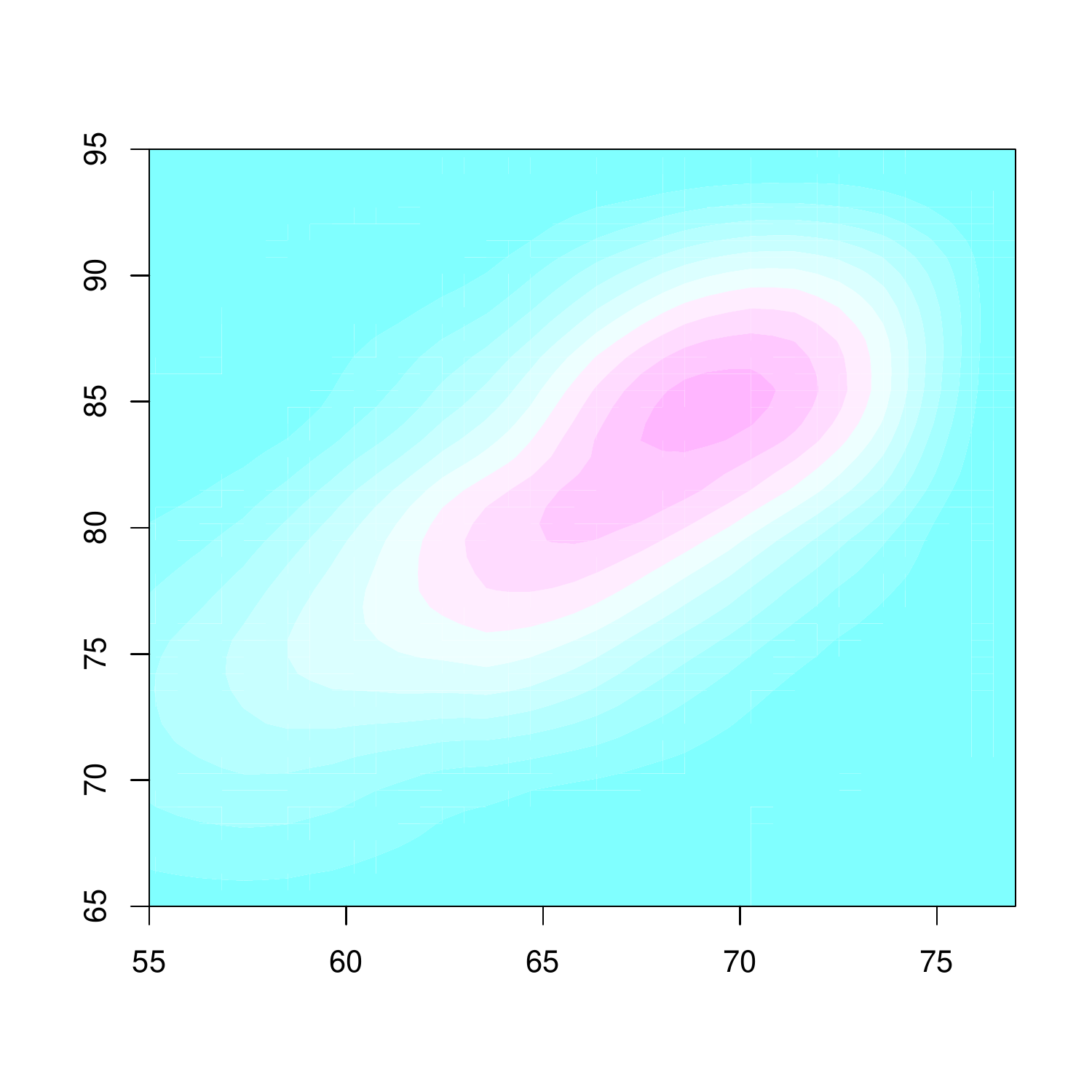}};
	 \node[above left = -1.4cm and -3.7cm of p22] (t22) {observed target};
	 \node[below = -0.2cm and 0cm of t22] (tt22) {(2019)};
	 & \node (p32) {\includegraphics[scale=0.33]{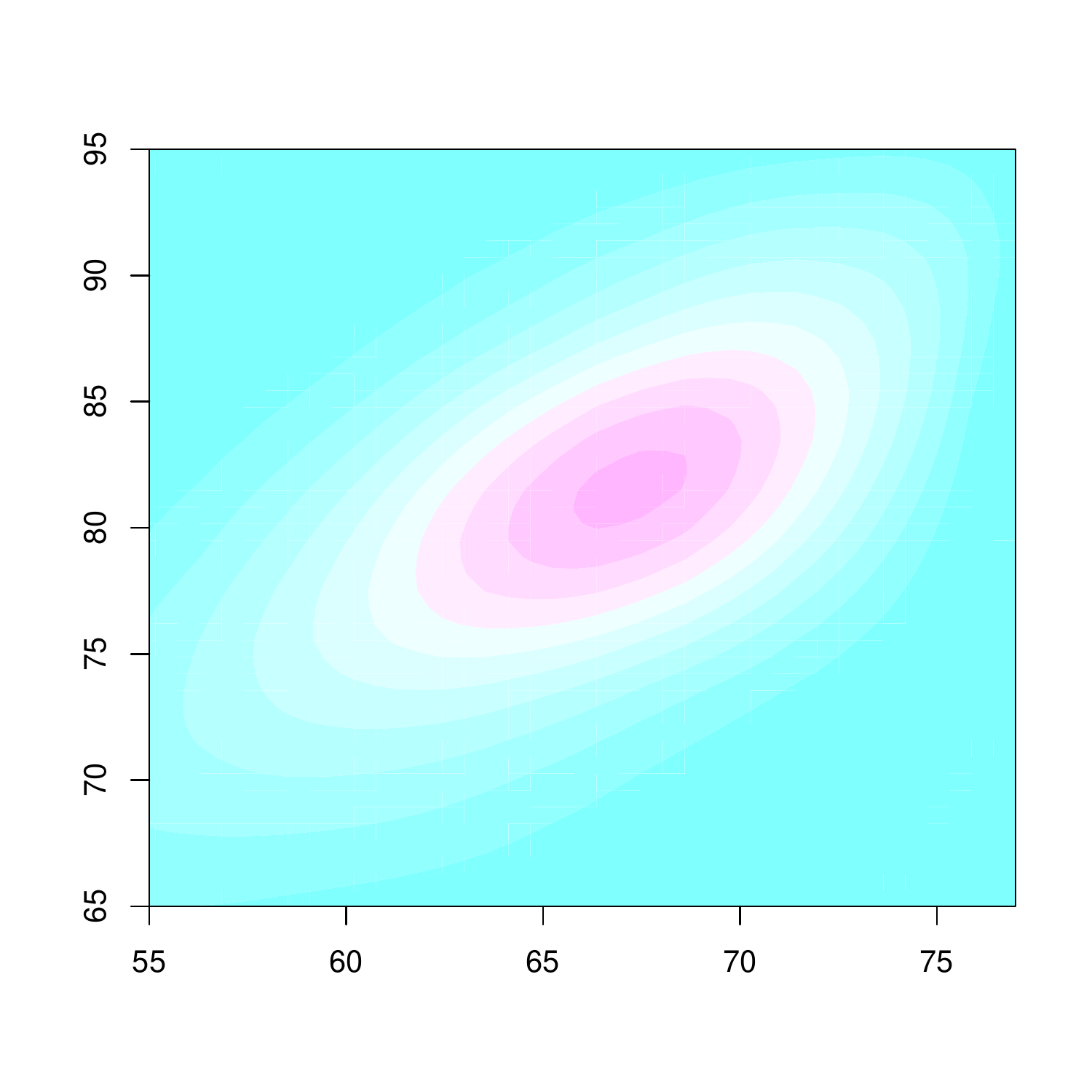}};
	 \node[above left = -1.4cm and -1.9cm of p32] (t31) {SAR};
	 & \node (p33) {\includegraphics[scale=0.33]{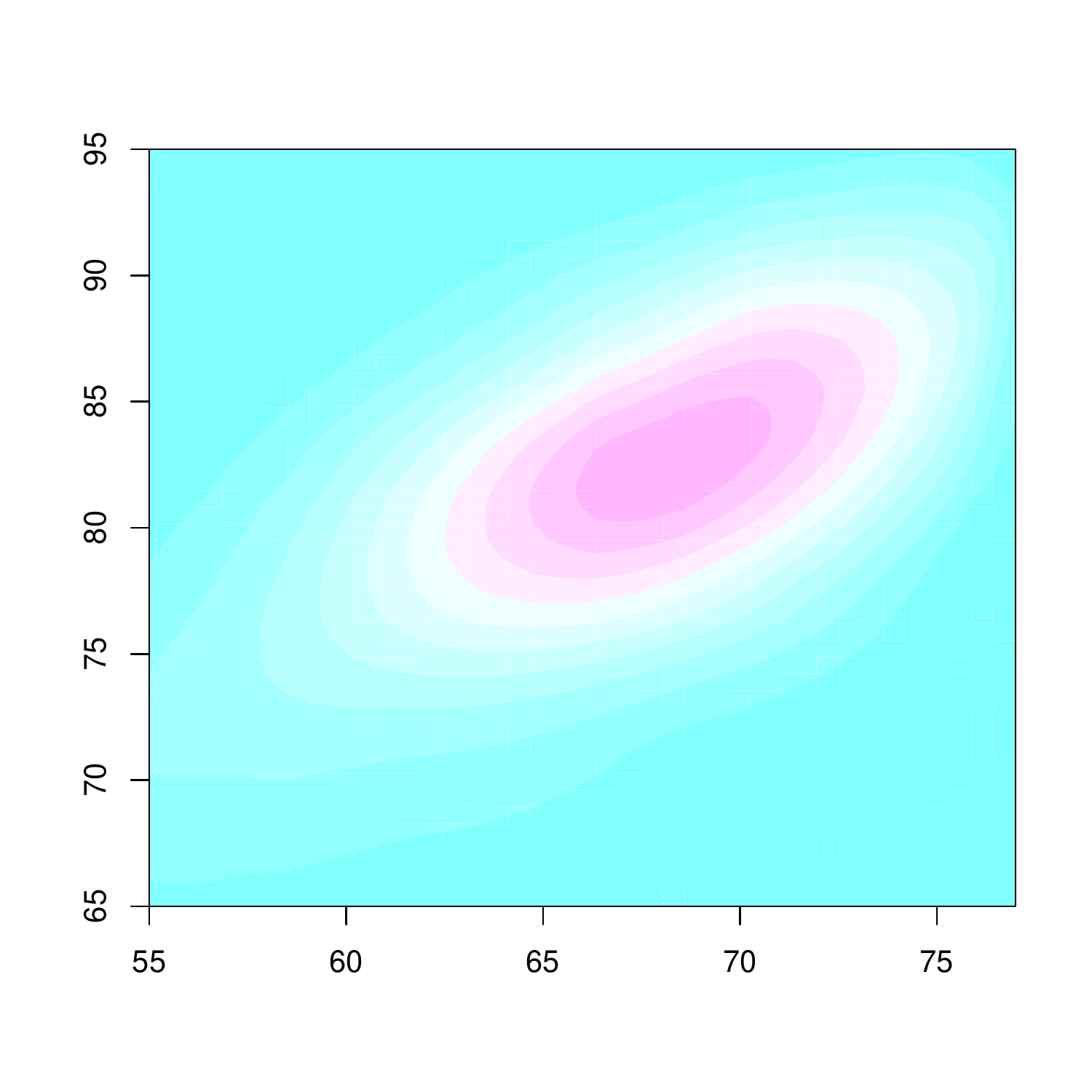}}; 
	 \node[above left = -1.4cm and -2.2cm of p33] (t32) {DSAR};
	 \\
};

\node[rotate=270,  below right=  -0.6cm and 0.7cm of p32] (legend) {\includegraphics[scale=0.25]{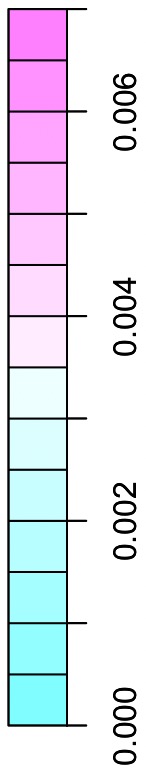}} ;
\node[above= -0.8cm and 0cm  of m] (title) {JFK International Airport} ;
\end{tikzpicture}
    \caption{Contour plots of observed and predicted two-dimensional density functions for the distributional time series of temperatures as recorded at JFK. The top six panels show  the observed density functions in the training set. The bottom left panels show the observed distribution for  2019 (left);  the predicted density using SAR (middle), with Fisher-Rao distance between predicted and observed of 0.147; and  the predicted density using DSAR, with Fisher-Rao distance 0.186.}
    \label{fig:jfk}
\end{figure}

\begin{figure}
    \centering
    \begin{tikzpicture}
    \matrix (m) [row sep = -3em, column sep = - 1.5em]{   
	 \node (p11) {\includegraphics[scale=0.7]{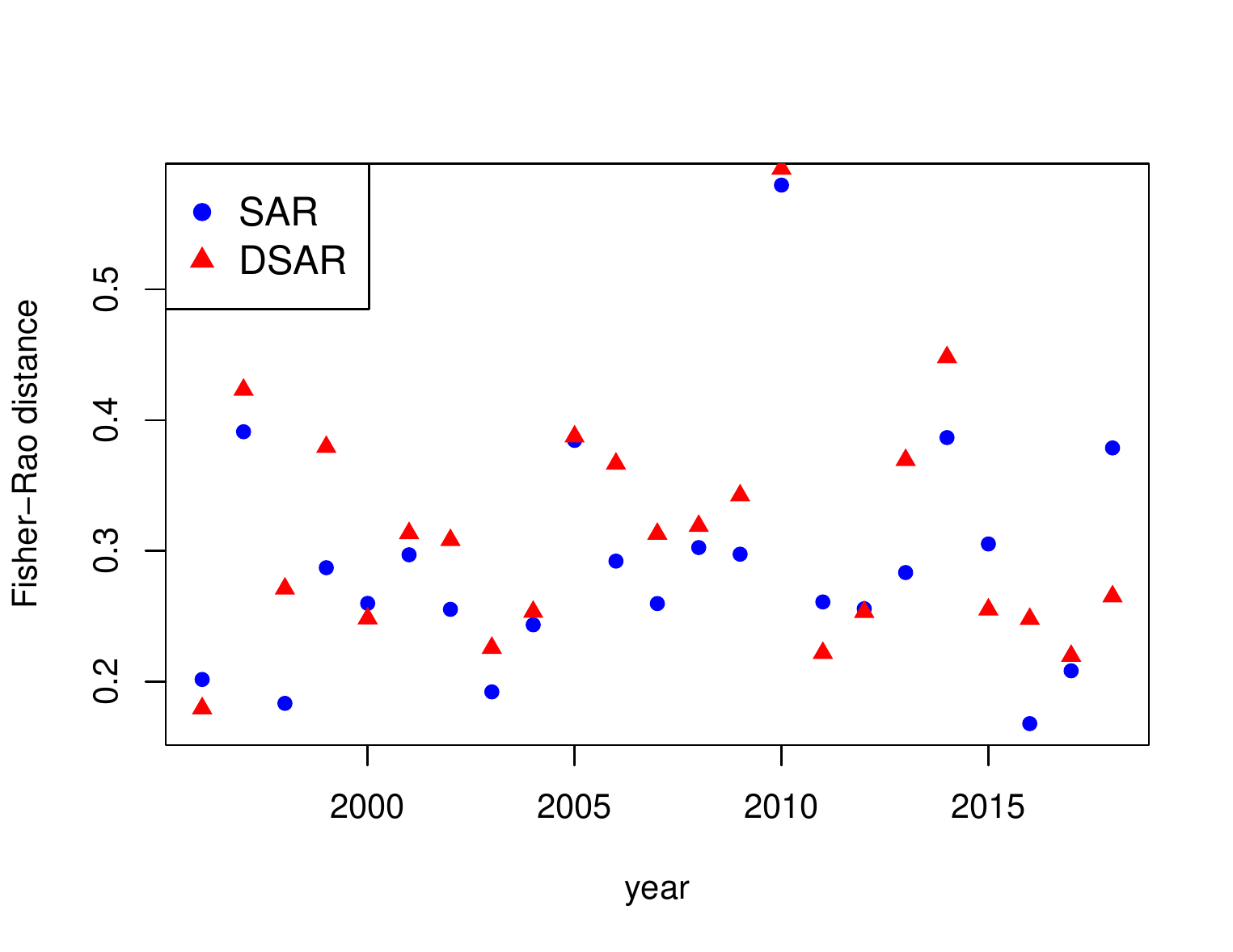}};
	 \node[above = -1.5cm and 0cm of p11] (t11) {LAX}; \\
	 \node (p12) {\includegraphics[scale=0.7]{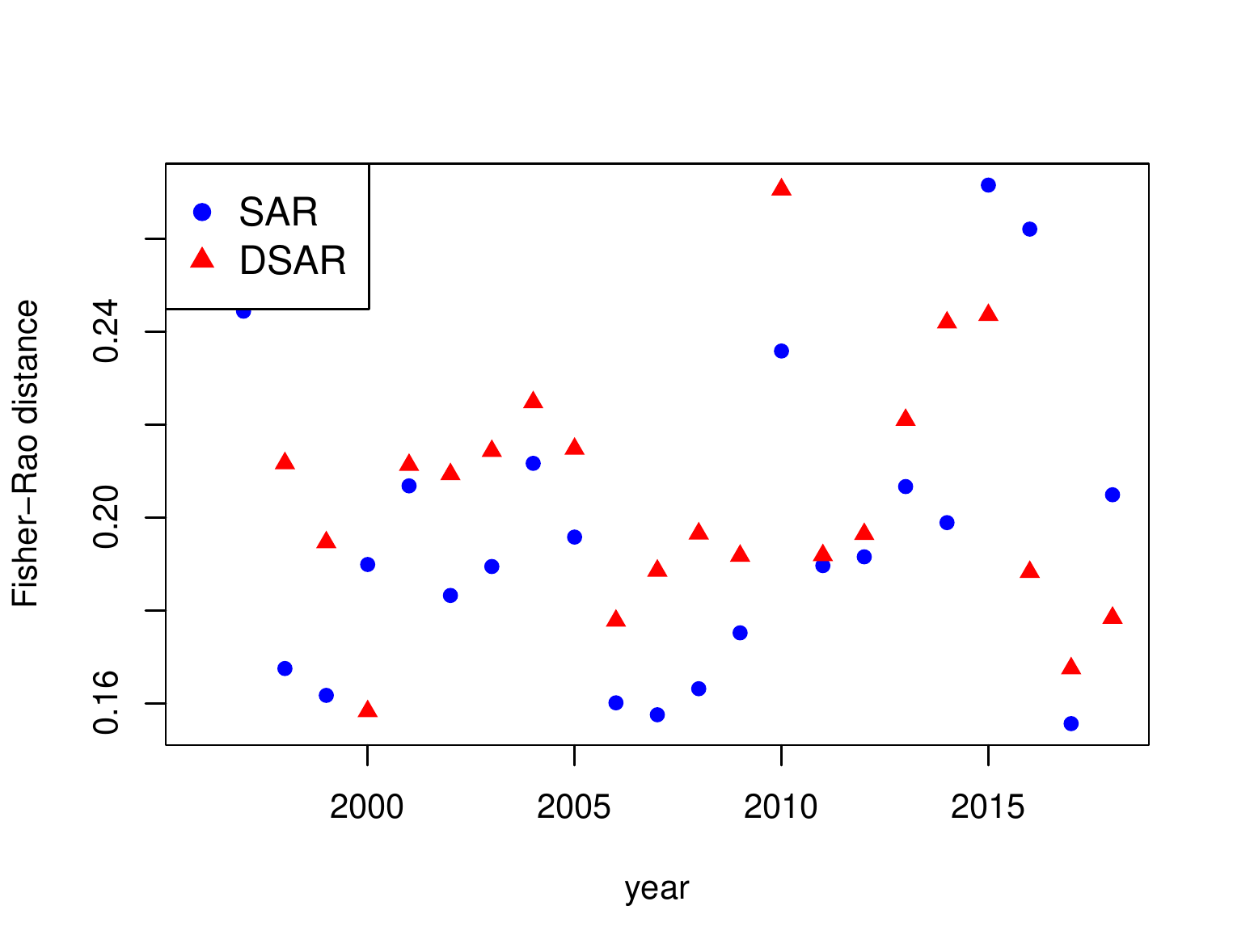}}; 
	 \node[above = -1.5cm and 0cm of p12] (t12) {JFK};
	 \\ 
    };
    \end{tikzpicture}
    \caption{Fisher-Rao distances between observed and fitted densities for each  year for distributional time series of two-dimensional temperature distributions.}
    \label{fig:fr}
\end{figure}

In a preprocessing step we obtained estimates of the bivariate density functions $f_t$ based on samples (\ref{sample}).  A quick and fast smoother that adjusts for boundaries is  histogram smoothing, which we implemented with histogram bins of size $50 \time 50$  and then applied the  R package ``fdapace" \citep{fdapace} for the smoothing step, where the bandwidth are set to be $(\max_i z_{t,i} -  \min_i z_{t,i})/5$ and $(\max_i w_{t,i} -  \min_i w_{t,i})/5$, then adjusting the results so that the estimated densities integrate to 1.  We thus obtained  30 bivariate density functions for the years from 1990 to 2019, some of which are shown as contour plots in the top six panels of  Figure \ref{fig:lax} and \ref{fig:jfk} for  Los Angeles International Airport
(LAX) and John F. Kennedy International Airport (JFK) respectively. We used the observed density for 2019 to illustrate the predictions obtained with SAR and DSAR using only the data before 2019 to construct the prediction.  The predicted densities are shown as contour plots at the bottom of Figure \ref{fig:lax} and \ref{fig:jfk}, where we chose the order $p=5$  for both SAR and DSAR.  We conclude  from both the contour plots and the Fisher-Rao distances that SAR works better than DSAR for this prediction, which is not surprising as the temperature distributions for JFK show much less year-to-year variation compared to those at LAX.  In addition, we plotted the FR distances between the observed and the fitted densities across time in Figure \ref{fig:fr}. There is no obvious trend, indicating a basic level of stationarity.  Interestingly, there is an obvious outlier for LAX for  2012, a year with the highest temperature on record (113 $^{\circ}$F) since 1921. \vs

\noindent { \sf 5.2 \quad Energy data}

\no Data on the sources of energy expressed as fractions or percentages for electricity generation across  the entire U.S. are available at \url{https://www.eia.gov/electricity/data/state/} and constitute a compositional time series. For our analysis we  consider three energy sources: (i)  Coal or Petroleum; (ii) Natural gas; (iii) Nuclear and Renewables. Sources (i) are known to produce the highest amounts of CO$_2$ and health damaging air pollutants per Watt generated, while sources (ii) are cleaner but still produce sizeable amounts of CO$_2$.  Sources  (iii) do not produce damaging gases while used for energy production but may have some residual risks such as nuclear energy production. Here we consider the compositional time series consisting of the annual proportions of energy generated from  sources (i)-(iii), which thus has three components. 

\begin{figure}
    \centering
     \begin{tikzpicture}
\matrix[
matrix of nodes, row sep = -8em,
nodes={
	anchor=center
}
]{  
	 \includegraphics[scale=0.9]{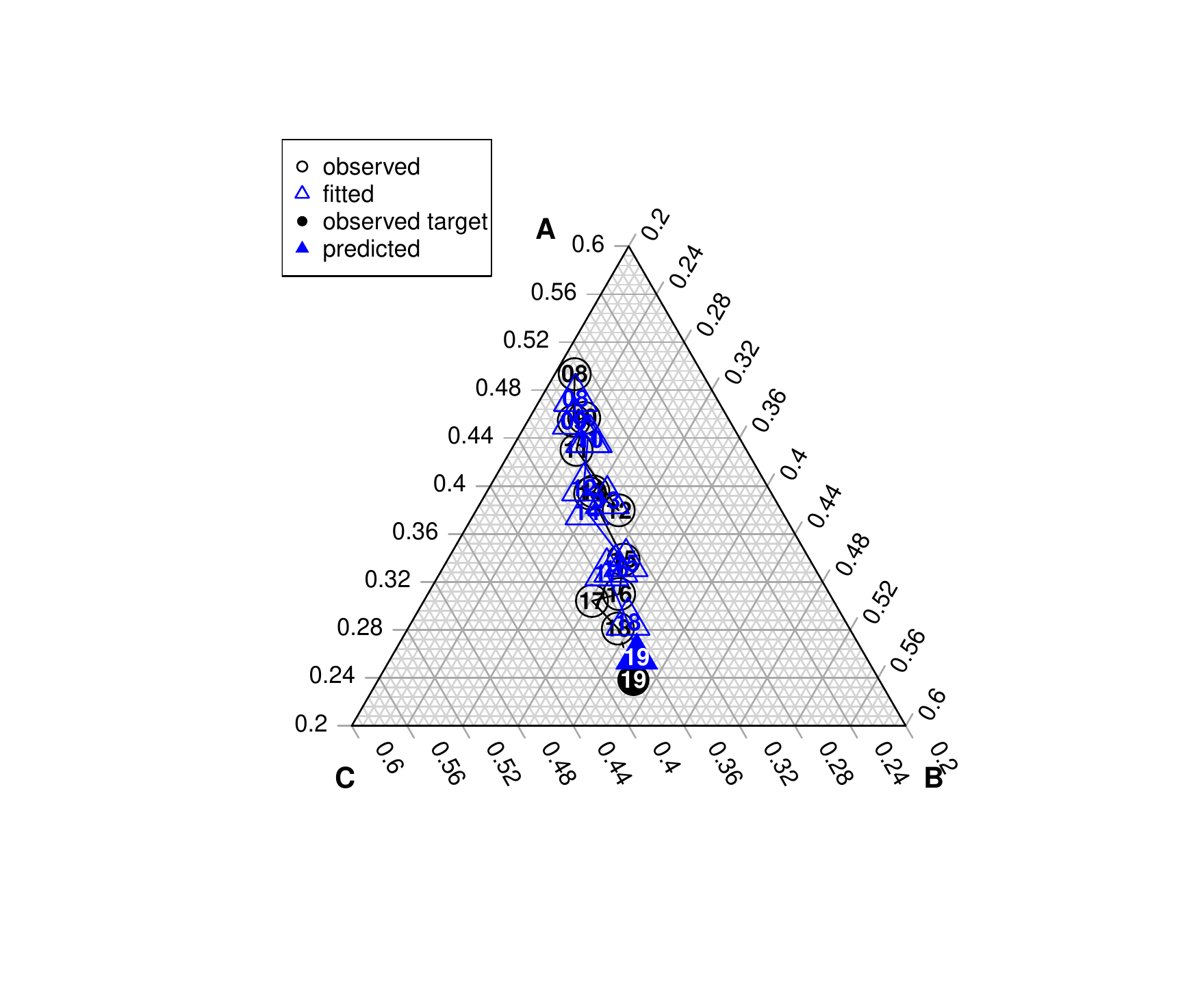} \\  \includegraphics[scale=0.52]{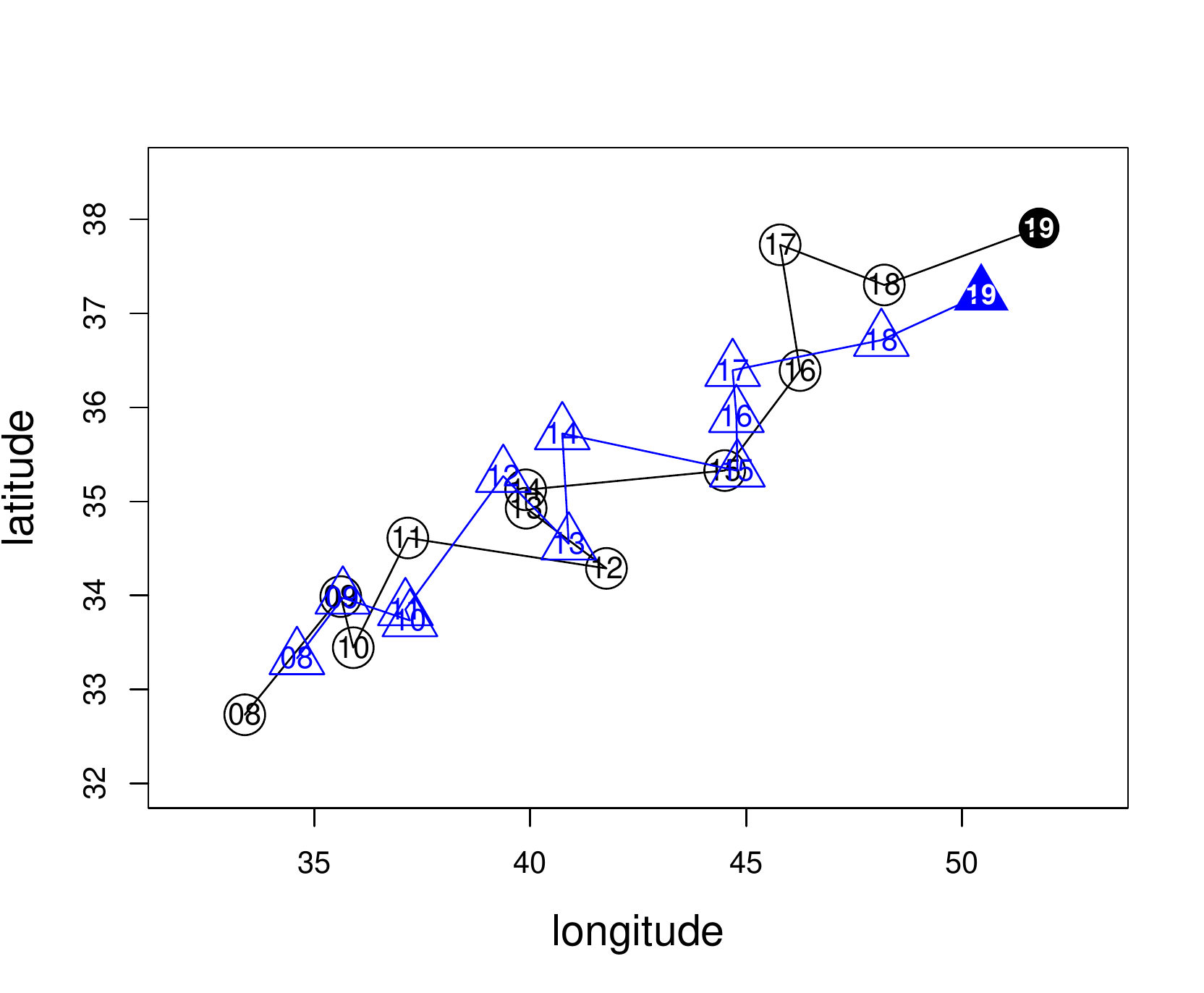} \\  
};
\end{tikzpicture}
    \caption{Observed (circles), fitted (triangles) and prediction target (19, circle is observed and triangle is predicted)  for the US energy sources compositional time series data, when fitting  model DSAR.  The numbers $ 08, \cdots, 09, 10, \cdots, 19 $ indicate the years from $2008$ to $2019$. Top panel:  Ternary plot reflecting the compositional nature of the data; here the corner A represents coal or petroleum; B represents natural gas; and C represents nuclear and renewables. Bottom panel: The compositional time series and predictions shown in spherical coordinates. The Fisher-Rao distance between predicted and observed compositions for 2019 is 0.0223. }
    \label{fig:energy}
\end{figure}

The data are available for the years $t=2005, 2006, \cdots, 2019$ and we denote the resulting time series by  $(U_t, V_t, W_t)$, where 
$U_t, V_t, W_t  \ge 0$ and  $U_t+V_t+W_t=1$ for all $t$. We then obtain the spherical time series $x_t = (\sqrt{U_t}, \sqrt{V_t}, \sqrt{W_t}) \in \mathcal{S}^{2}$. The data $\{ x_t \}_{t=2005}^{2018}$ are used as training set to fit SAR and DSAR models and we aim  to predict the proportions of the energy sources for the year 2019.
The observed compositions from 2005 to 2018 and the observed, fitted and predicted compositions for 2019 
are shown in Figure \ref{fig:energy} and illustrated  with two types of graphical representations for compositional data.  A ternary plot is in the top panel and spherical plot in the bottom panel,  where for the latter we  plotted the longitude and latitude of each point $x_t \in \mathcal{S}^2$.

Both plots show a strong trend over the years and the ternary plot indicates  that the proportion of energy generated from source (iii) is continuously increasing each year. Correspondingly, the proportion of energy from coal or petroleum is continuously decreasing. The  trend indicates some degree of non-stationarity of $x_t$, while no trend seems to be present when considering the  annual increments that correspond to the spherical rotations from one year to the next.  It thus appears that  the differences $\{ x_{t+1} \ominus x_{t} \} $ are sufficiently stationary. Consequently,  we applied model DSAR, for  order $p=2$. Figure \ref{fig:energy} indicates that  DSAR not only fits the observed data quite well but also produces  a reasonable prediction for the energy mix in the   year 2019.\vs

\bc {\bf \sf \Large 6.\quad Discussion}\sm \ec \rs

While both compositional and distributional time series can be represented as spherical time series, such time series also arise for directional data \cp{mard:14}. Vector time series may also be represented with a spherical component if  one is primarily interested in the directions of the vectors over time and less in their length, via polar coordinates.  All of this adds to the  
motivation to study spherical time series, while  at the same time, there is little  methodology available at this time. 
In this paper we attempt to address this dearth of methodology
by 
developing  an autoregressive modeling approach. We propose to represent rotation operators on spheres by skew-symmetric  operators that  can be viewed as elements of a Hilbert space so that linear operations become available. Other approaches may also be possible but they have not yet been developed. Our goal is to provide a first modeling approach for this situation as a baseline with which future approaches can be compared. 

It is of course possible to use different metrics for both compositional and distributional time series.    For compositional data, a classical alternative is the Aitchison geometry \cp{aitc:86}, which also  has been extended to distributional data \cp{hron:16}. 
However, in applications to compositional data this approach does not work  if some of the component fractions are zero and then requires arbitrary adjustments, and it also requires the arbitrary selection of a baseline component; the spherical approach does not face these difficulties  \cp{scea:14}. 

For distributional time series an obvious alternative is to consider the space of distributions equipped with the   Wasserstein metric \cp{vill:03} that is connected with  optimal transport. When adopting this metric,  the time series is not spherical and needs to be modeled in the Wasserstein manifold, where one can use tangent bundles \cp{mull:21:4,zhan:21} or an intrinsic optimal transport approach \cp{mull:21:3}.  However when dealing with the Wasserstein space  for multivariate distributions one faces major hurdles in both theory and computation. In contrast  the Fisher-Rao metric that we consider here allows seamless extensions to any dimension. When the distributions are unknown, they need to be estimated and density estimation in higher dimensions is subject to the curse of dimensionality.  This can be counteracted by assuming that the number of data from which each of the densities is  estimated is large.

Further  in-depth comparisons of the various possible approaches to distributional and 
compositional time series will need to await future research. Beyond  these two signature applications, autoregressive models for spherical time series provide a useful tool for directional time series and other situations where one has a natural representation of data on a finite- or infinite-dimensional sphere. Another area of future research will be the development of other time series approaches for such data that extend autoregressive models to more complex models for time series such as GARCH models or to the frequency domain. 

Finally, the spherical regression models that we have proposed here are also applicable 
for the case of a multiple regression in a non-time series context, for situations where both predictors and responses are spheres. In this case  
one has $n$ i.i.d. pairs $(X_{i1},\dots,X_{im}, Y_i) \in \S$ and aims  to model and  obtain fits for the regression relation $E(Y|X_{1},\dots,X_{m})$. To our knowledge, such multiple spherical regression models have not been studied yet.\vs

{
	\bibliographystyle{agsm}
	\bibliography{sphere,1-1-21}
}

\end{document}